\documentclass[%
 aip,
 amsmath,amssymb,
 reprint,%
]{revtex4-1}

\usepackage{graphicx}
\usepackage{dcolumn}
\usepackage{bm}

\usepackage[utf8]{inputenc}
\usepackage[T1]{fontenc}
\usepackage{mathptmx}
\usepackage{makecell}
\usepackage{xcolor}
\usepackage{ulem}
\usepackage{lipsum}
\begin{document}

\preprint{AIP/123-QED}

\title{Aggregation dynamics of methacrylate binary and ternary biomimetic polymers in solution\\}

\author{Garima Rani}
 \email{grani@imsc.res.in}
 \affiliation{The Institute of Mathematical Sciences, C.I.T. Campus, Taramani, Chennai 600113}
 \affiliation{Homi Bhabha National Institute, Training School Complex, Anushakti Nagar, Mumbai 400094, India}
\author{Kenichi Kuroda}%
 \email{kkuroda@umich.edu}
\affiliation{ 
Department of Biologic and Materials Sciences, School of Dentistry, University of Michigan, Ann Arbor, Michigan 48109, United States
}%

\author{Satyavani Vemparala}
\email{vani@imsc.res.in}
 \affiliation{The Institute of Mathematical Sciences, C.I.T. Campus, Taramani, Chennai 600113}
 \affiliation{Homi Bhabha National Institute, Training School Complex, Anushakti Nagar, Mumbai 400094, India}


\date{\today}

\begin{abstract}
Using detailed atomistic simulations, we explore the conformational landscape of aggregates formed by biomimetic antimicrobial (AM) binary methacrylate copolymers, with hydrophobic and charged functional groups and the role of inclusion of polar functional groups on such aggregate morphologies. The effect of sequence of the constituent functional groups on aggregate conformation is also studied by considering random and block sequences along the polymer backbone. Our results suggest that block binary copolymers form large spherical aggregates with effective shielding of hydrophobic groups by charged groups.  In contrast, random binary copolymers tend to form more bundle-like structures with exposed hydrophobic groups. The strong aggregation of binary polymers are driven primarily by attractive interactions between hydrophobic groups. However, replacing some of the hydrophobic groups with overall charge neutral polar groups weakens the aggregate considerably, leading to increased conformational fluctuations and formation of loose-packed, open aggregates, particularly in the case of random ternary polymers. Interaction energy calculations strongly suggest that the role of inclusion of polar groups is two-fold: (1) to reduce possible strong local concentration of hydrophobic groups and "smear" the overall hydrophobicity along the polymer backbone to increase the solubility of the polymers (2) to compensate the loss of attractive hydrophobic interactions by forming attractive electrostatic interactions with the charged groups and contribute to aggregation formation, albeit weak. Given that most of the naturally occurring AM peptides have contributions from all the three functional groups, this study elucidates the functionally tuneable role of inclusion of polar groups in the way antimicrobial agents interact with each other in solution phase, which can eventually dictate their partitioning behavior into bacterial and mammalian membranes.

\end{abstract}

\maketitle

\section{\label{sec:level1}INTRODUCTION}

Biomimetic antimicrobial (AM) polymers have been a focus of research in recent times due to the possibility of their usage as novel therapeutic agents against infectious pathogens \cite{Locock14,Takahashi17,ageitos2017antimicrobial,REVmcphee05,REVboman95}. Such polymers are designed based on the basic functional themes present in the large class of naturally occurring AM peptides  \cite{hwang1998structure,zhang2016antimicrobial,pasupuleti2012antimicrobial,REVnguyen11}. The two predominant functional groups identified in the AM peptides are hydrophobic and charged groups, with the cationic functionality of the AMPs providing selective binding to anionic lipids of the bacterial cell membrane by electrostatic attractions, whereas their hydrophobic residues facilitate insertion into the non-polar membrane core. These have been replicated in the design of AM polymers as well. The distribution of hydrophobic and charged groups along the polymer backbone broadly results in the emergence of two distinct classes of polymers: rigid and flexible antimicrobial AM polymers \cite{Tang06,Som2008}. In either class of antimicrobial polymers facial amphiphilicity, in which the charged and hydrophobic groups are segregated along the backbone has been emerging as one of the key determinants of the antimicrobial efficacy \cite{Tew2002,Som2008,Palermo18,zasloff02}. In AM polymers with rigid backbones, such a facially amphiphilic distribution of charged and hydrophobic is intrinsic in the design \cite{aryl2,Firat04,phen1,phen2}. However, the rigidity of the backbone in addition to their in-built facially amphiphilic nature, might prevent them from acquiring novel structures which could facilitate new modes of insertion into bacterial membrane. To address these issues, there is also substantial focus in designing polymers with flexible backbone to mimic the structural features and functions of antimicrobial peptides, due to their ease of preparation, lower cost and potent antimicrobial activity against a broad spectrum of bacteria \cite{Kenichi05,palermobook,Palermo18}. Our previous studies based on methacrylates, combining experiments and MD simulations, revealed that flexible copolymers which lack built-in amphiphilicity can adopt facially amphiphilic conformations at the water-membrane interface \cite{Ivanov06, Palermo12, baul14}. 

Several studies have highlighted the role of key structural parameters such as the percentage of hydrophobic groups, ionic groups and molecular weight in modulating the biological activity of AM polymers in order to achieve high antimicrobial efficacy, while minimizing toxicity to mammalian cells \cite{Firat04, Kenichi05,Kenawy07,Palermo18,mowery2009structure}. Experimental studies have also shown that it is essential to understand the conformation of AM polymers in solution along with their aggregation potential which can affect their eventual interactions with bacterial membranes \cite{Oda18,sato2009macromolecular,nakashima2006aggregation,Oda11}. For instance, in one of the earliest simulation study of the  solution-based conformational landscape of designed biomimetic polymers based on methacrylate moieties\cite{Ivanov06}, it was shown that both the length and sequence (arrangement of different functional groups along the polymer backbone) play crucial role in the polymers attaining unique configurations in solution. Experimental studies on comparing the role of sequence of charged and hydrophobic groups on the antimicrobial as well as hemolytic properties of the polymers have indicated that block copolymers can more effectively encapsulate the hydrophobic groups as compared to the random distribution of the same \cite{Oda11,Oda18}. These findings strongly suggest that not only the percentage of charged and hydrophobic groups, but the sequence in which these groups are ordered along the polymer backbone plays a significant role in driving the polymer-polymer interactions in solutions. Individual polymer conformations in solutions can potentially affect their aggregation behaviour and can eventually affect their interactions with membrane and the consequent membrane disruption modes, which are at the heart of the antimicrobial mechanism of both natural and biomimetic polymers. 

Most of the studies have focused on inclusion of only charged and hydrophobic groups as primary constituents of designed biomimetic polymers \cite{Uppu16, Takahashi17, Mitra15,Yang18}. However, in such random binary copolymers it is very difficult to optimize the monomer composition appropriately for potent antibacterial activity and reasonable selectivity. Highly hydrophobic polymers form strong aggregates in the solution which might lead to two undesirable consequences: (1) the strong solution aggregate may preclude partition of the individual AM polymers into the bacterial membrane (2) such aggregates may bind to human cell membranes due to non-specific hydrophobic interactions, causing undesired toxicity to humans. On the other hand, low hydrophobic content in the AM polymer may lead to stabilising the structures in the solution and decrease their ability to penetrate into the bacterial membranes. Hence a requisite amount of hydrophobicity is needed to increase the local concentration of the AM polymers for efficient interaction and eventual partitioning of the AM polymers into the membrane which requires a fine tuning of not only the hydrophobic content of polymers, but spreading of the same along the polymer backbone.  In this context, more recent works \cite{wang08, wang2012importance, Chakraborty14, Yang14, Gao11, Mortazavian18} have been focusing in inclusion of new functional groups like polar groups, in addition to charged and hydrophobic groups, which also comes closer to the functional group distribution in naturally occurring AM peptides. This opens up not only new paradigms of designing more effective AM polymers with more nuanced mechanisms, but also helps in delineating the specific contributions of different functional groups of such polymers. Chakraborty et al. using ternary nylon-3 copolymers reported that replacing hydrophobic or charged groups or both by hydroxyl residues can significantly reduce the hemolytic activity of the copolymer compared to their binary counterpart with only hydrophobic and charged subunits \cite{Chakraborty14}. Other studies with charged and hydrophilic groups also exhibited significantly lower hemolytic activity, compared to the methacrylate random polymers with charged and hydrophobic moieties \cite{Yang14}. A polymer brush based implant coating consisting of covalently grafted hydrophilic polymer chain conjugated with an optimized series of AMPs was found to act effectively against biofilm formation and showed good antimicrobial activity, both in vivo and in vitro \cite{Gao11}. Mortazavian et al. used ternary copolymer system to decouple the effects of charged and hydrophobic functional groups \cite{Mortazavian18}. Their study demonstrated that the plausible role played by polar group in the polymer is to reduce the sequential domains of strong hydrophobic monomers, which can modulate polymer chain insertion into bacterial and human cell membranes.

In this work, our aim is to understand the role played by inclusion of polar groups and specific sequence of functional groups along the polymer backbone in the aggregation dynamics of methacrylate copolymers in solution phase using detailed atomistic simulations. In this context, multiple polymers with compositions of  binary (only charged and hydrophobic groups) and ternary (charged, hydrophobic and polar groups)  and with random and block arrangements of the functional groups are studied in solution phase to understand the individual as well as aggregate morphologies. Our studies show that the role of inclusion of polar groups is to redistribute the effective hydrophobicity along the polymer backbone and result in weaker, yet functionally relevant conformations compared to polymers with strong hydrophobic content, due to larger conformational fluctuations. The sequence of the functional groups also crucially dictates the morphology of the aggregates with block copolymers exhibiting more compact structures. Such loosely packed "weak aggregates" can play a crucial functional role in antibacterial action, by making the polymer dissociation from the solution aggregate easier which might lead to the favourable partitioning of the polymers into the bacterial membrane and consequently membrane disruption.

\section{\label{sec:MD}MODEL AND SIMULATION METHODS}

\begin{figure}[ht]
 \centering
\includegraphics[width=\columnwidth]{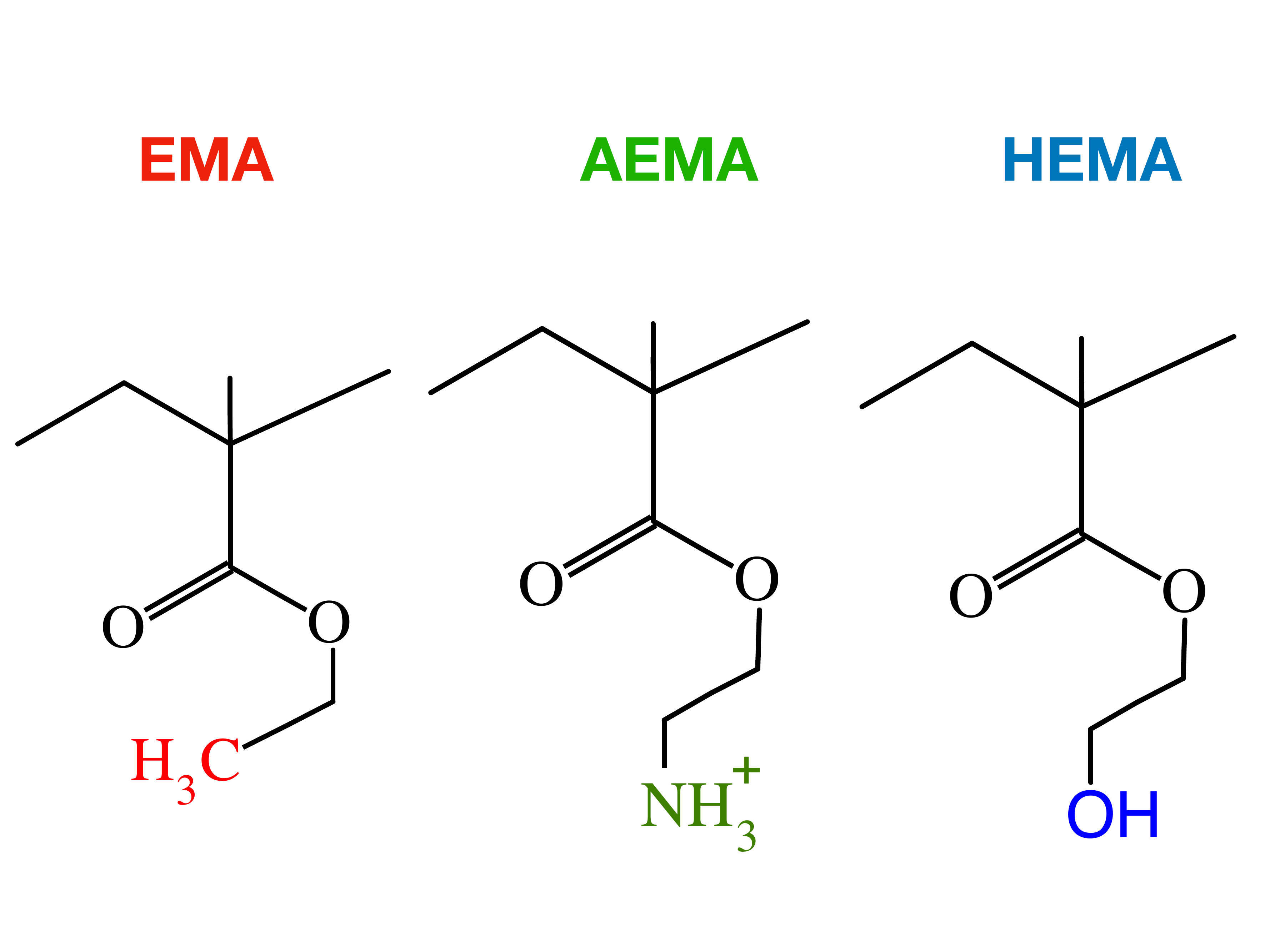}
\caption{\footnotesize Chemical structures of EMA, AEMA and HEMA groups considered in the model polymers.}
 \label{fig:chem}
 \end{figure}
Atomistic MD simulations with explicit water and ions were performed on two classes of model biomimetic copolymers- binary and ternary copolymers. Ternary methacrylate random copolymers ("model T"), consisting of charged ammonium (amino-ethyl methacrylate: AEMA), hydrophobic alkyl (ethyl methacrylate: EMA) and neutral hydroxyl (hydroxyl methacrylate: HEMA) groups as shown in Fig.~\ref{fig:chem}, are modelled with degree of polymerization (DP) = 19. The chosen composition of ternary copolymer  (AEMA-6, EMA-8 and HEMA-5) has been shown to be optimal for antimicrobial activity and also shows significantly reduced hemolytic activity\cite{Mortazavian18}. To understand the role played by sequence of functional groups, we performed simulations involving ternary polymers which have a block arrangement of functional groups, consisting of same $\%$ of groups as model T but having blocks of sequences 6(AEMA)-5(HEMA)-8(EMA) along the polymer backbone ("model TB"). Further, to highlight the role of inclusion of polar hydroxyl functional groups, control simulations without them involving only binary compositions of charged (AEMA) and hydrophobic (EMA) monomer units in random ("model B") and block ("model BB") configurations, were also performed. The degree of polymerization was kept same as for ternary polymers and the number of monomers per chain was taken to be 6 (AEMA) and 13 (EMA). Details of the composition of various subunits in the model polymers are summarized in TABLE I and the detailed chemical structures of all the four model polymers are shown in Figure 1 of the Supplementary Information. It is important to note that for all the model polymers, the number of charged functional groups are fixed to be 6 per polymer in agreement with Mortazavian et al \cite{Mortazavian18}, where it was found that despite differences in EMA composition, the minimum inhibitory concentration (MIC) values of all polymers start to level off at $\sim 30\%$ of AEMA and charged groups above 30 $\%$ do not increase the antimicrobial activity of the polymer. This proportion of charged group is also comparable to net positive charge of most natural AMPs\cite{zasloff02}.

\begin{table}
\resizebox{\columnwidth}{!}{%
\begin{tabular}{|*{7}{c|}}\hline
\hline
\makebox[3.7em]{\makecell{Model\\copolymers}} & \makebox[10em]{\makecell{Polymer$\%$\\(AEMA, HEMA, EMA)}}&\makebox[1.8em]{DP}&\makebox[2.8em]{\makecell{AEMA\\ No.}} & \makebox [2.8em]{\makecell{HEMA\\ No.}} & \makebox[2.5em]{\makecell{EMA\\ No.}}\\
  &     &  &   &  &\\
\hline
Ternary (T, TB) & (31.58, 26.32, 42.10) & 19 & +6 & 5 & 8 \\
Binary (B, BB)  &  (31.58,0,68.42) & 19 & +6  & 0  & 13\\
\hline
\hline
\end{tabular}}%
\Large{\caption{Composition of AEMA, HEMA and EMA subunits in the model polymers.}}\label{tab:table1}
\end{table}

A single polymer chain for each system was built in an extended conformation and solvated with TIP3P \cite{jorgTIP83} water. The total charge of a single polymer chain in each of the system was $+6e$, and requisite salt ions were added to both neutralize the systems and to maintain 150 mM salt concentration to mimic physiological conditions in each case. All simulations were performed with the NAMD 2.9 simulation package \cite{Phillips05} and the simulations were performed in a rectangular 3-D box with periodic boundary condition applied. Each system of polymers was first energy minimized with the conjugate gradient method.  Single polymer chain simulations were done with 2 fs timestep and performed in the (isothermal - isobaric) NPT ensemble for 2 ns.  Next, systems of aggregated polymers were built by selecting 10 random conformations from the single polymer simulation trajectory. This group of 10 initially dispersed polymers for each model (T, TB, B and BB) were placed in a TIP3P water with a minimum distance of 12 \r{A} between any polymer atom and the box side, under periodic boundary conditions and 150 mM of NaCl salt concentration was maintained to neutralize the system. Initially, the systems were energy minimized with conjugate gradient method and equilibrated  with 2 fs timestep  for 5 ns of NVT simulation to stabilize the system and all the subsequent runs for atleast 150 ns were done in the NPT ensemble (All the polymer systems and simulation details are listed in TABLE 1 of Supplementary Information). All the MD simulations were performed at constant temperature maintained at 305 K with Langevin dynamics at a collision frequency of 5 ps$^{-1}$, and a pressure of 1 atm was maintained through Langevin piston \cite{Martyna94,Feller95}. Electrostatic interactions were calculated by the Particle Mesh Ewald method\cite{Essmann95} and the cut-off for non-bonded interactions was set to 12 \r{A}, with smoothing starting from 10 \r{A}. The parameter values for the polymers were adopted from the CHARMM force field \cite{karplus98} and previous simulations \cite{Ivanov06,Palermo12,baul14}. The largest aggregate in each run is extracted for further analysis to understand the properties of the aggregation dynamics of  model polymers in solution. Three  independent simulations were performed for each model polymer with different initial configurations of dispersed polymers. The visualization was done using software VMD \cite{vmd96} and the analysis for the polymer aggregates was done by taking average over multiple independent runs, unless otherwise stated, using TCL scripting language which is embedded with VMD. 

\begin{table*}
\resizebox{6.5in}{!}{%
\centering
\begin{tabular}{|*{8}{c|}}\hline
\hline
\makebox[10em]{Model}& \makebox[10em]{Arrangement} &\makebox[10em]{Group sequence} &\makebox[10em]{Largest Aggr. size, $N_{agg}$}\\
  &  &  &   S1, S2, S3\\
\hline
   Binary copolymers  & (AEMA-HEMA)   & &  \\
  B & R  & E-E-A-E-E-A-E-E-A-E-E-A-E-E-E-A-E-E-A  & 4, 4, 5 \\
 BB  &  B & A-A-A-A-A-A-E-E-E-E-E-E-E-E-E-E-E-E-E & 6, 10, 5 \\
 \hline
 Ternary copolymers & (AEMA-HEMA-EMA)  & &  \\
 T  &  R & E-H-A-E-H-A-H-E-A-E-E-A-E-H-H-A-E-E-A  & 4, 5, 4  \\
 TB  & B & A-A-A-A-A-A-H-H-H-H-H-E-E-E-E-E-E-E-E & 5, 7, 4\\
\hline
\hline
\end{tabular}}
\caption{Summary of the models polymers and their largest aggregation size, denoted $N_{agg}$, at the end of 150 ns NPT simulation runs. In each case, initial configuration consists of 10 randomly dispersed polymers in a box of water and ions. Here, R and B denote  random and block arrangement, groups in the polymer are represented as E (hydrophobic EMA), A (cationic AEMA) and H (Polar HEMA). }\label{tab:table2}
\end{table*}

\section{\label{sec:result}RESULTS}
\subsection{Morphology of the aggregates}

\begin{figure}[h]
 \centering
\includegraphics[width=\columnwidth]{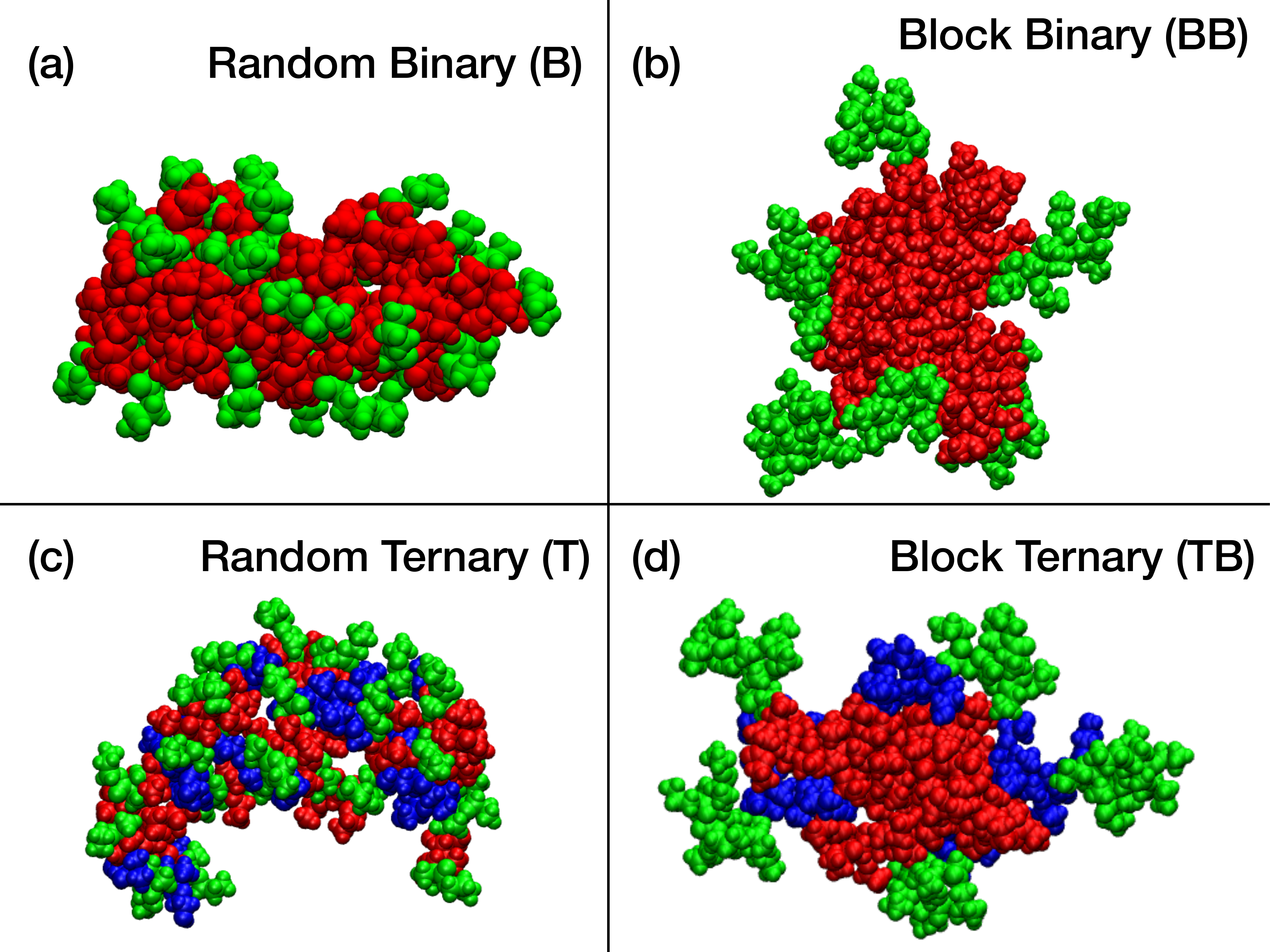}
\caption{\footnotesize Representative conformation of the aggregates in the VDW representation  with different side chains, hydrophobic (EMA, red), amino (AEMA, green), and polar (HEMA, blue), at the end of one of the 150 ns long MD simulations in NPT ensemble for binary copolymers (a,b) and ternary copolymers (c,d). The number of polymers in a typical aggregate, $N_{agg}$ for the snapshots shown are 4 (model B), 6 (model BB), 4 (model T) and 5 (model TB) (see Supplementary Information TABLE~\ref{tab:table1} for more details).}
 \label{fig:aggr}
 \end{figure}
 
In this Section, we study the role of inclusion of polar functional group and also the importance of understanding how sequence of functional groups in the AM polymers affect the morphology of the aggregates in solution. All the polymers with and without polar groups and with random and block arrangements of functional groups form aggregates in the solution, albeit with different morphologies and intra-aggregate interactions similar to the previous experimental results \cite{Oda18}. Illustrative snapshots of the aggregates in solution, at the end of one of the representative simulation for all the four models are shown in Fig~\ref{fig:aggr}. Each model was independently simulated three times to confirm that the final aggregation morphology is independent of the starting state of the polymers and to ensure reproducibility of results (see Supplementary Information Figure 2 for more details). In Table \ref{tab:table2}, we present the sizes of the largest stable aggregate formed for each model polymer within the simulation time, with the number of polymers in this aggregate denoted as $N_{agg}$. Visual inspection of the trajectories indicate that the evolution of aggregates for different polymer models is markedly different- the aggregates formed from random binary (model B) polymers form relatively compact assemblies displaying significant entanglement of the constituent polymers while the random ternary polymer (model T), with partial replacement of hydrophobic EMA groups by hydroxyl HEMA groups, form somewhat more open and comparatively less entangled aggregates. The binary block copolymer (model BB) form larger sized aggregates, showing a star-like micellar conformation with a dense hydrophobic core surrounded by charged subunits, exposed to water solution. The ternary block (model TB) also form large aggregates compared to its random counterpart, though in this case the HEMA group remains clustered between EMA and AEMA monomers. Our aim now is twofold, to analyse (a) the effect of inclusion of polar groups (binary vs ternary) and (b) the effect of variation in the sequence of the AEMA, EMA and HEMA groups along polymer backbone (block vs random) on the shapes and sizes of the aggregates formed. \\

\begin{figure*}[ht]
\centering
\includegraphics[width=6in]{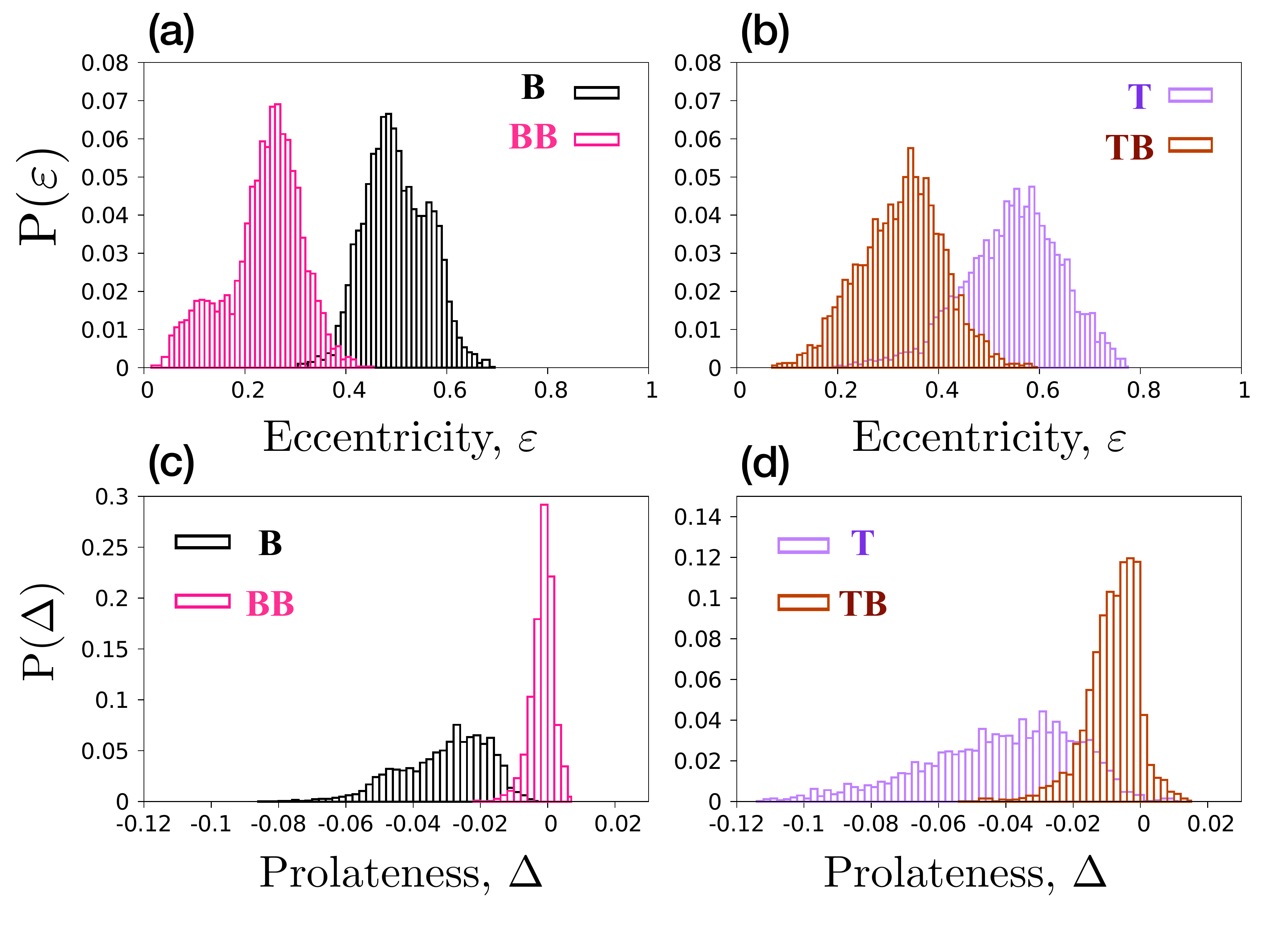}
\caption{\footnotesize Probability distribution of the eccentricity, $\varepsilon$ and prolateness, $\Delta$ measured for the aggregates formed from the models (B, BB, T and TB) from the start of aggregation formation till the end of the 150 ns simulation runs, averaged over all the initial conditions.}
 \label{fig:ecc}
 \end{figure*}

\noindent\textbf{Shape Parameters -} We consider two shape parameters \cite{ecc17,pro2006}, eccentricity ($\varepsilon$) and prolateness ($\Delta$) to describe the morphologies of aggregates for the four model polymer systems, both of which are calculated from the moment of inertia tensor ($I$) of the aggregate as:
\begin{align}
\varepsilon &= 1-\frac{I_{\text{min}}}{I_{\text{avg}}}\,, \\
\Delta &= \frac{\displaystyle\prod_{i=1}^{3} (I_{i} - I_\text{avg})}{I_\text{avg}^3}\,,
\end{align}
where $I_{\text{min}}$  and $I_{\text{avg}}$  denote the minimum  and the average values of the principal moments of inertia ($I_{1}$, $I_{2}$ and $I_{3}$). For perfectly spherical objects $\varepsilon, \Delta=0$ and $\varepsilon=1$ for linear objects while negative and positive values of $\Delta$ correspond to oblate and prolate ellipsoid shapes.

The results for probability distribution of these shape parameters of the aggregates formed from the model polymers over the time from the start of stable aggregation formation till the end of the simulation run are shown in Figure~\ref{fig:ecc}. Data in Figures ~\ref{fig:ecc}(a) and \ref{fig:ecc}(b) shows that the mean and width of the distribution of eccentricity ($\varepsilon$) values of ternary polymer aggregates are larger than those of binary polymer aggregates (B and BB models), suggesting a more extended structure and also the fluctuating dynamic conformations in case of the model T and TB aggregates. The average value of $\varepsilon$ is smaller for the aggregates formed from block polymers (BB, TB) compared to random polymers (B, T) and this strongly shows that aggregates formed from block polymers are more compact compared to the aggregates formed from random polymers.  The probability distribution of the prolateness ($\Delta$) for all the four models is shown in Figures ~\ref{fig:ecc}(c) and \ref{fig:ecc}(d) and the data confirms that the block polymers indeed form more spherical and compact structures ($\Delta \approx 0$) but the random polymers with negative values of $\Delta$ quantify the oblate shape of their aggregates. The data also suggests that the conformational dynamics of the aggregates formed by the random ternary polymers is much higher than that of random binary polymers. The fluctuations in the values of both $\varepsilon$ and $\Delta$ can be seen from the time evolution over the last 20 ns of simulations, as shown in Supplementary Information Figure 3(a).The shape parameters ($\varepsilon$ and $\Delta$) thus demonstrate the following - the shape parameter values for aggregates in which the hydrophobic content is large and which have no additional polar groups (as in model B and model BB aggregates) are narrowly distributed, hence displaying less variability in their conformations. On the other hand, the aggregates of the polymers having polar HEMA groups (model T and model TB aggregates) show significantly larger variance in the values of the shape parameters. This underlines the role of the polar HEMA groups in inducing conformational fluctuations in these aggregates. However, the $\Delta$ and $\varepsilon$ data does suggest that on the average, the block polymer aggregates (model BB and model TB) have a closer to spherical morphology as compared to their respective random polymer aggregates counterparts. This is likely due to the presence of hydrophobic groups distributed across the backbone of the random polymers, which then leads to the gluing of the polymer chains in a more bundle-like conformation.

\begin{figure}[h]
\centering
\includegraphics[width=\columnwidth]{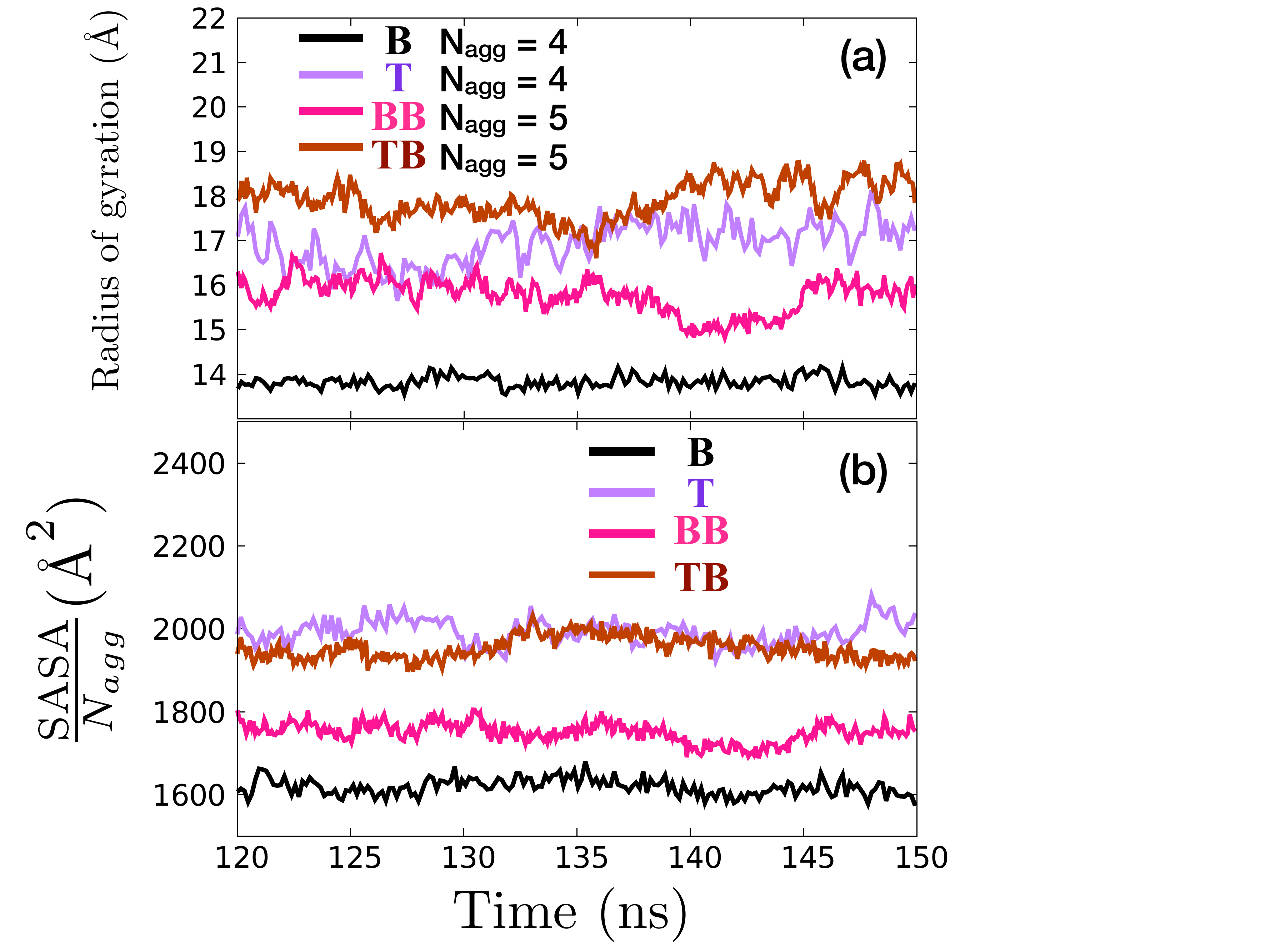}
\caption{\footnotesize (a) Radius of gyration for random polymers (models B, T) having aggregate size $N_{agg} = 4$ and block polymers (models BB, TB) having aggregate size $N_{agg} = 5$ is plotted as a function of time,  (b) Average SASA$/N_{agg}$ is plotted as a function of time.}
 \label{fig:rg-sasa}
 \end{figure}

\noindent\textbf{Size measures -} One of the measures of aggregate dimensions and stability is the radius of gyration \cite{rg2008}, defined as the root mean square distance of the constituents of the aggregate from its center of mass. We compute the radius of gyration $R_g$ of the aggregates as a function of time over the last 30 ns of the simulations for random polymer aggregates with $N_{agg}=4$ and block aggregates with $N_{agg}=5$, as shown in Fig.~\ref{fig:rg-sasa}(a). We observe that in both cases, the ternary aggregates exhibit a higher $R_g$ value, highlighting the role of polar HEMA groups in inhibiting formation of strong, well packed compact aggregates. Interestingly, the $R_g$ value of model T aggregate is comparable to the $R_g$ value of model TB aggregate, even though the aggregate size $N_{agg}$ is larger for model TB aggregate (4 vs 5). This in particular shows that the random ternary (model T) polymer aggregate has the least compact structure amongst the polymer aggregates considered here, forming a loose and open aggregate conformation.  

\begin{figure}[h]
\centering
\includegraphics[width=\columnwidth]{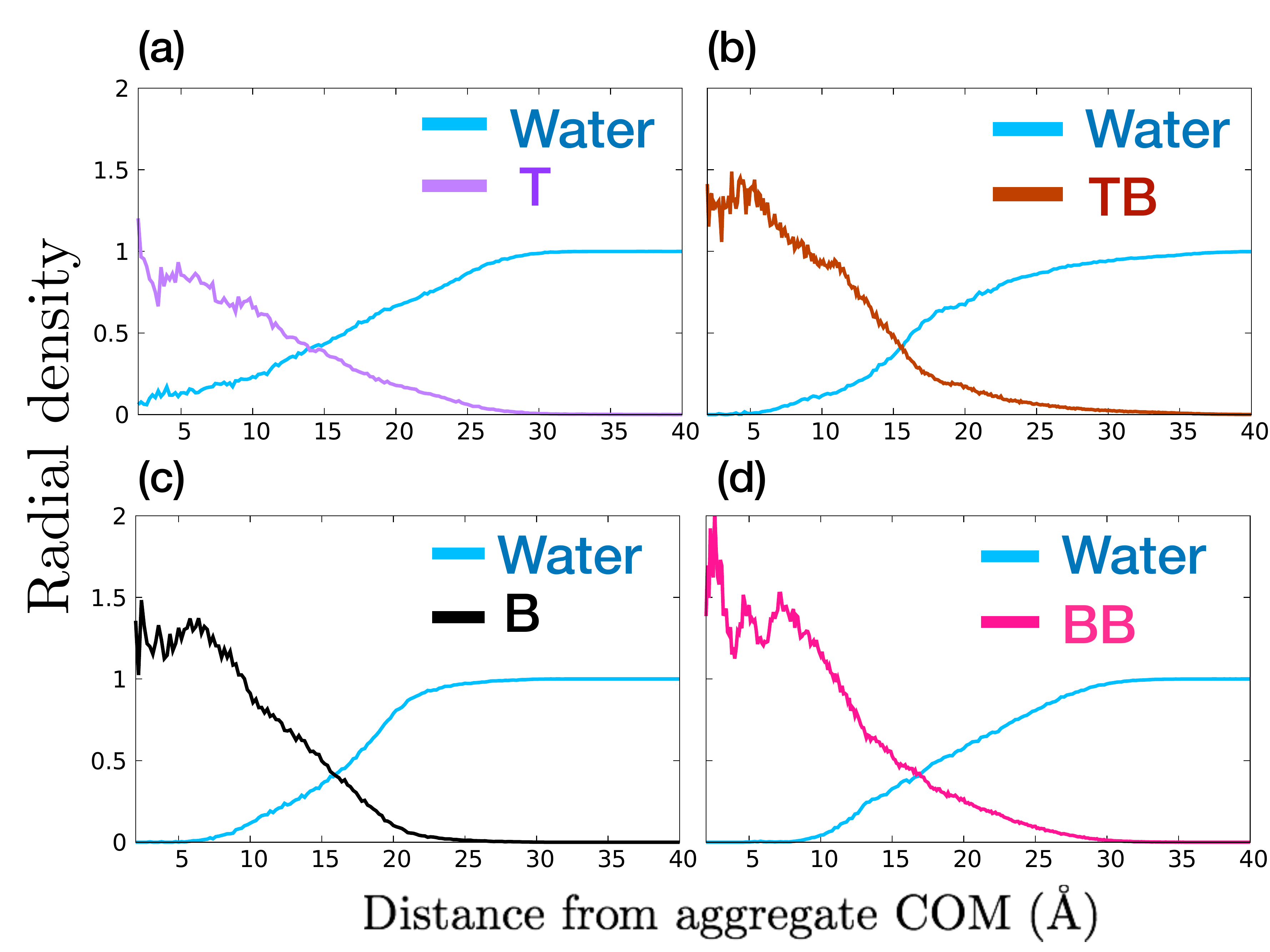}
\caption{\footnotesize Radial density profiles of the aggregate and water molecules plotted as a function of distance from the COM of the aggregate for (a) model T, (b) model TB, (c) model B, (d) model BB. The analysis is done over the last 30 ns (120-150 ns) of one of the simulation runs (S1).}
 \label{fig:ag-w}
 \end{figure}

\noindent\textbf{Solvent Accessible Surface Area and aggregate-water interaction -} Structural properties of the aggregates can also be inferred by measuring the Solvent Accessible Surface Area (SASA) \cite{sasa71} of the aggregate and average SASA of the aggregate per polymer, since $N_{agg}$ varies with the polymer model considered, is shown in Figure~\ref{fig:rg-sasa}(b). The total SASA values per aggregate is also calculated (Supplementary Figure 4) and was calculated with a spherical probe of radius 1.4 \r{A}.  The higher values of total SASA for block polymer aggregates as compared to random polymer aggregates is due to micellar structure that are formed with predominant exposure of both charged (AEMA) and polar (EMA) groups to solvent and efficient sequestration of hydrophobic EMA groups at the core. The random arrangement of functional groups along the polymer backbone frustrates such efficient partitioning of the polymers in an aggregate resulting in lower SASA values. This in particular is true for binary random polymer (model B) aggregate, which has the lowest SASA value of all the polymer aggregate models. The importance of sequential arrangement of the functional groups along the polymer backbone can be seen in significant differences of overall SASA values for for binary random and binary block polymer aggregates. We have also measure SASA/polymer and those results (Figure~\ref{fig:rg-sasa}(b)) show that for ternary polymers, the SASA values are not appreciably different between the random and ternary polymer aggregates. To quantify the nature of interaction between the aggregates and bulk water in general and to estimate the exposure of hydrophobic core to water in particular, the radial density of water is calculated as a function of distance from the aggregate center of mass by counting water molecules in 0.1\r{A} of shell width around the center of mass and compared with the radial density of the aggregate itself (Figure~\ref{fig:ag-w}). We observe that in case of model T aggregate (fig.~\ref{fig:ag-w}(a)), the density of water is already non-zero very close to the COM of the aggregate but it slowly increases to reach the maximum bulk density far from the center of mass. This underscores that the core of the model T aggregate is weakened with significant exposure to water (a plausible consequence of the presence of hydrophilic HEMA groups which favour interactions with water) and further, that it has a loosely packed, somewhat open structure, as was also deduced by our previous analysis. On the other hand, in case of the model TB aggregate (fig.~\ref{fig:ag-w}(b)), non-zero water density only starts to appear at $\sim 5$ \r{A} from the center of mass, indicating that it has a comparatively strong hydrophobic core. However, in general we observe that non-zero water density starts earlier and shows a comparatively slower rise to reach bulk water density in case of ternary aggregates as compared to the binary aggregates. This suggests that the presence of HEMA groups results in formation of loosely packed aggregates which adopt more extended conformations. For model B polymer system (fig.~\ref{fig:ag-w}(c)), we observe that non-zero water density starts away from the center of mass and it increases sharply to reach the maximum bulk density, confirming its compact aggregate structure with relatively high hydrophobic density near the center of mass, again showing good agreement with our previous analysis. In case of the block binary model BB (fig.~\ref{fig:ag-w}(d)), water density is zero for a long distance from the center the aggregate. This, in conjunction with the radial density profiles of EMA and AEMA groups (Supplementary figure 5) and our previous analysis, clearly shows that in this case, large sized aggregates are formed whose hydrophobic core is densely packed with EMA residues, is relatively impermeable to water and is surrounded by charged residues in a spherical micellar conformation (in case of model TB aggregate as well, the radial density profiles (Supplementary figure 5) exhibit that the charged residues primarily lie on the boundary of the aggregate surrounding the EMA and HEMA groups).

Our data allows us to make the following conclusions - block polymer systems tend to form larger aggregates as compared to random systems. Further, binary aggregates form well packed, compact aggregates while the block binary aggregate displays a spherical micellar conformation with a strong hydrophobic core which is almost impermeable to water, surrounded by charged groups, the random binary aggregate displays a more ellipsoidal shape. Finally, presence of polar HEMA groups leads to conformational fluctuations and in general results in formation of loosely packed, open aggregates, particularly in the case when the groups are randomly distributed along the polymer backbone. Such loosely packed "weak aggregates" can play a crucial role in antibacterial action. In these cases, polymer dissociation from the aggregate can occur, with polymer-lipid head groups of the bacterial membrane interactions becoming favourable over polymer-polymer interactions more easily, as compared to well packed, stable aggregates \cite{Horn12,baul14}. This can lead to polymer insertion into the bacterial membrane and thus, membrane disruption and eventual cell death.  The results also show that a high hydrophobic content can result in a strong aggregate and this can result in two undesirable results: their interactions with bacterial membranes may be ineffective or they can be highly toxic to mammalian membranes \cite{Tominaga10,Oda11,Oda18}. A possible role of the polar HEMA groups can be to modulate the intra aggregate interactions and affecting polymer-membrane interactions.  In the next section, we study how the inclusion of HEMA groups can affect the inter-polymer binding, to highlight its role in the partitioning of the polymers from the aggregate phase to the membrane phase.

\subsection{HEMA groups in aggregation dynamics - binary vs ternary}

\begin{figure}
\centering
\includegraphics[width=4.5in]{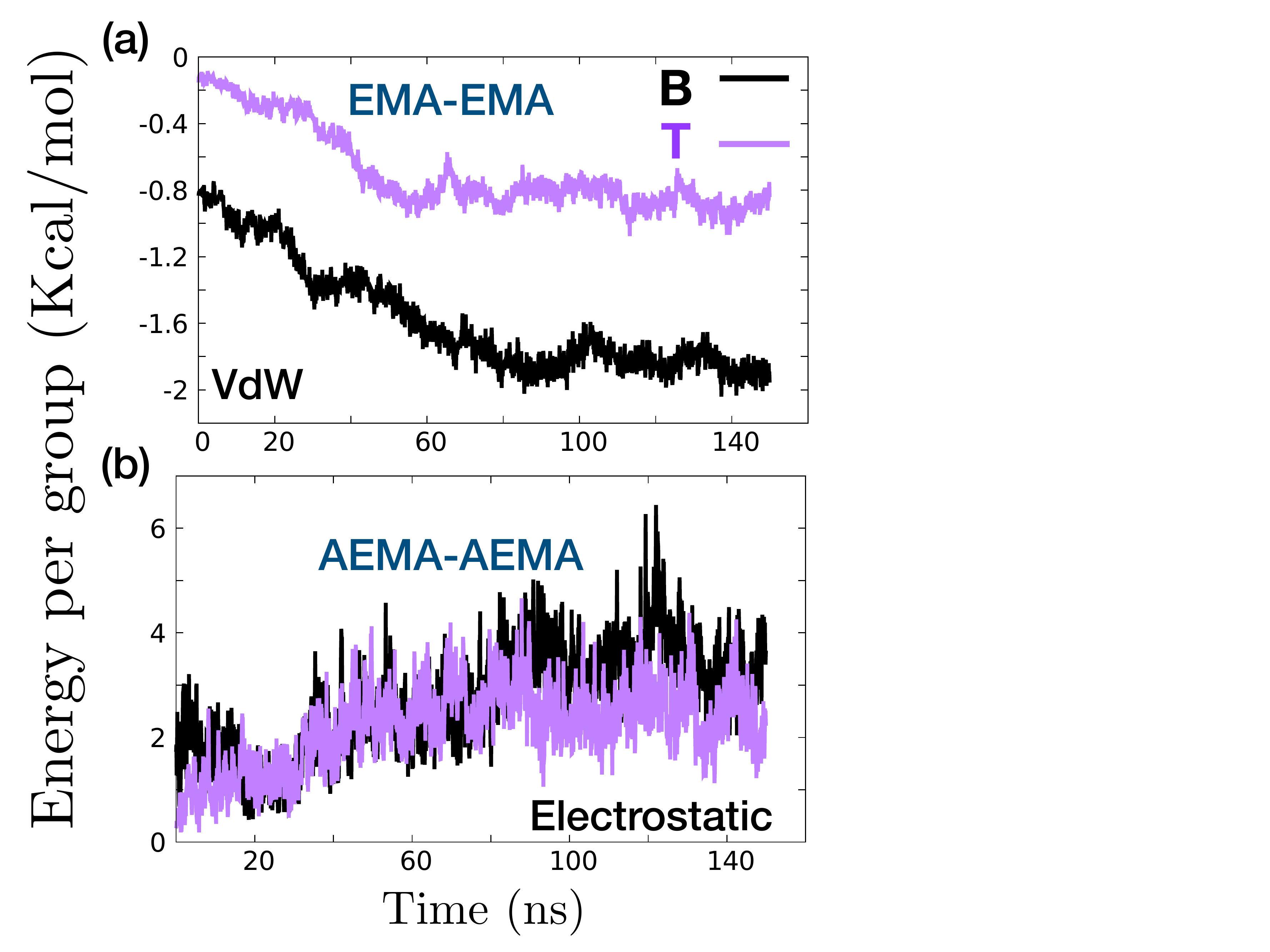}
\caption{\footnotesize Pair interaction energy per unit group plotted as a function of time. Model B is shown in black colour and model T is shown in purple colour. (a) Van der Waals energy per EMA for hydrophobic groups and (b) electrostatic energy per AEMA for charged side chain groups are shown for both the polymer aggregates.}
 \label{fig:energy1}
 \end{figure}

\begin{figure}[h]
 \centering
\includegraphics[width=\columnwidth]{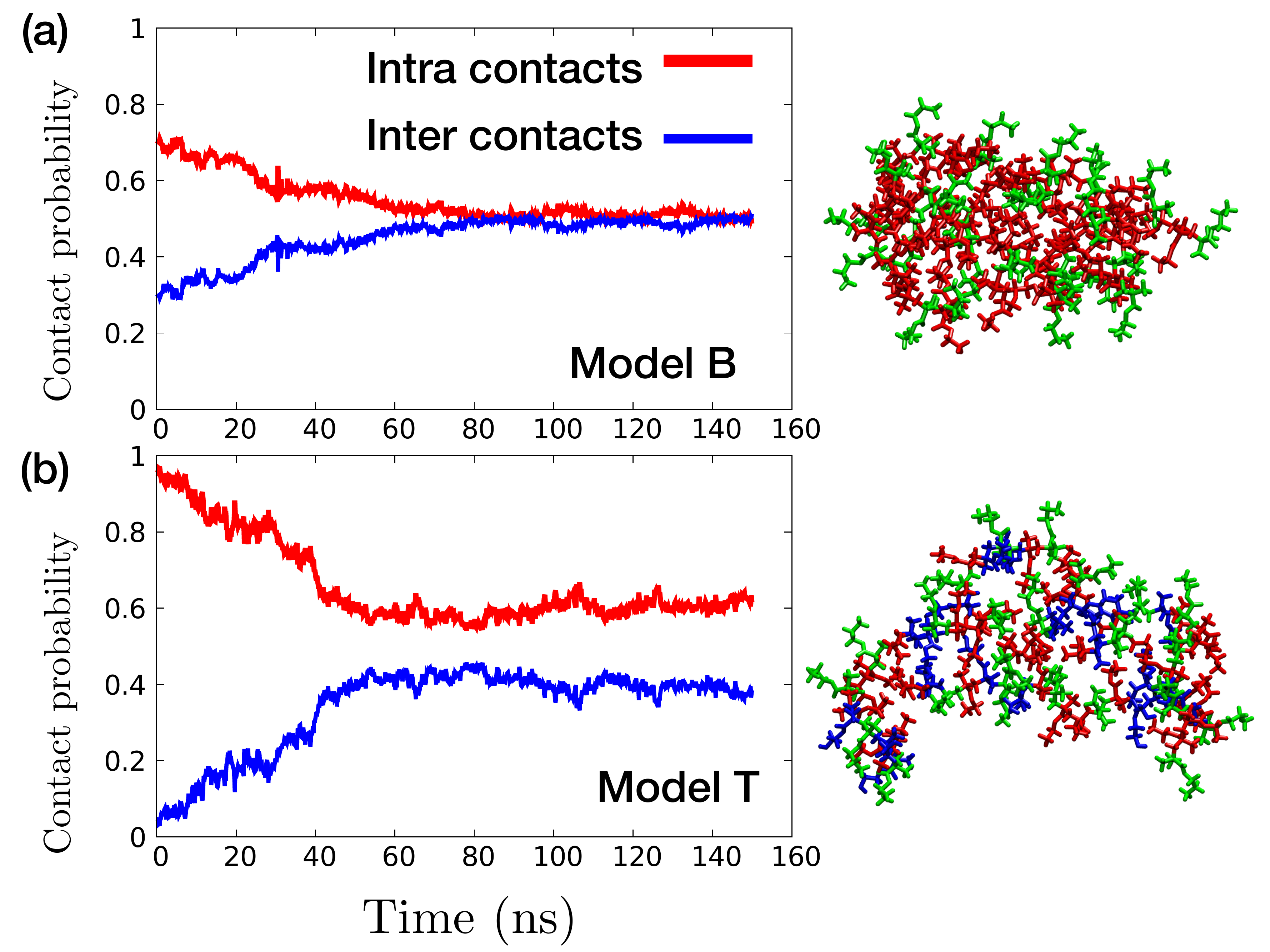}
\caption{ \footnotesize Average contact probability between hydrophobic EMA residues that belong to same polymer (intra-contacts) and different polymers (inter-contacts) plotted as a function of time for (a) model B, (b) model T. Representative snapshots for the binary and ternary random model aggregates are depicted to the right, with EMA, AEMA and HEMA groups represented in red, green and blue color respectively.}
\label{fig:contacts}
\end{figure}

In this section, our aim is to understand the role of the polar HEMA groups on aggregation dynamics, in particular the effect of replacing hydrophobic EMA groups by polar HEMA groups on the stability of aggregates formed, particularly in case of random aggregates (models B and T). In this context, we first note that the total van der Waals interaction energy is higher for the model B aggregate, as expected due to the presence of more hydrophobic EMA groups. However, the question then arises whether the same is true even after the effect of the presence of higher proportion of hydrophobic moieties is discounted. Our data in Figure~\ref{fig:energy1}(a) shows that indeed this is the case, with the the total van der Waals energy per EMA group remaining appreciably more attractive for the model B aggregates. This indicates that on the average, EMA-EMA interactions are strongly attractive in this case and thus, contribute considerably to the stability of the aggregate. On the other hand, for the model T aggregate, the interactions are much weaker due to the presence of polar HEMA groups, which clearly impede the formation of a strong aggregate. However, the electrostatic interactions between models B and T are not significantly different due to the presence of same number of charged AEMA groups in both the set of polymers (Figure~\ref{fig:energy1}(b)). Therefore, the more attractive van der Waals interactions in the model B aggregate is a clear indication that the hydrophobic EMA groups primarily drive the aggregation formation for model B polymers. This is further illustrated by considering the inter polymer and intra polymer contact probabilities of the EMA-EMA moieties, as shown in Fig. \ref{fig:contacts}. A contact between two EMA groups is said to exist if the two are within $7$ \r{A} of each other and is classified as an intra-contact if the EMA groups belong to the same polymer or as an inter-contact, otherwise.  We deduce from Fig. \ref{fig:contacts} that in case of the model B aggregate, as it evolves, the inter polymer contacts match the intra contacts, which strongly suggests that polymers in the aggregate are entangled with each other and consequently, it has a much more tightly interwoven structure, driven primarily by EMA-EMA interactions. However, this is manifestly not true for the model T aggregate, in which case the intra contacts always remain higher than the inter contacts, which indicates that the EMA-EMA interactions are not the primary driver of aggregation dynamics in this case.  It also suggests that the individual polymers in the model T aggregate are most likely in a self-compact structure with more intra-polymer contacts and fewer inter polymer contacts. This is further proof that the partitioning of the polymers from the aggregate will be significantly easier in model T polymer aggregates than for model B polymer aggregate.\\
 
\begin{figure}[h]
 \centering
\includegraphics[width=\columnwidth]{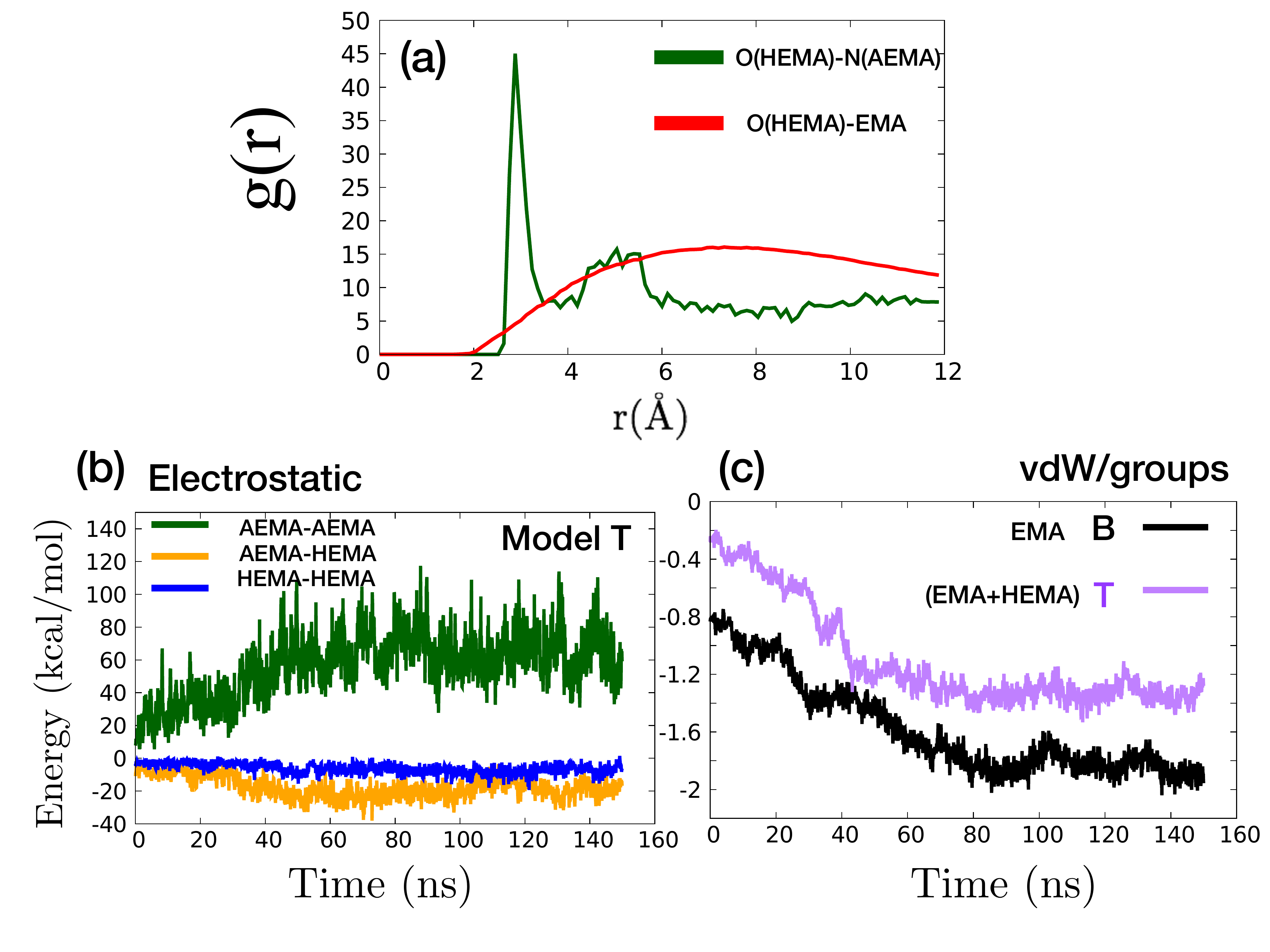}
\caption{ \footnotesize (a) Radial density function g(r) for model T, computed over the last 30 ns of one of the simulation run (S1). (b) Total electrostatic energy is plotted as a function of simulation time, computed between AEMA-HEMA, AEMA-AEMA and HEMA-HEMA groups of the model T aggregate. van der Waals energy per group is plotted as a function of simulation time, computed between EMA-EMA groups of model B aggregate and (EMA+HEMA)-(EMA+HEMA) groups of model T aggregate are calculated.}
\label{fig:gr}
\end{figure}
\noindent\textbf{Role of HEMA groups in ternary aggregation - } For the ternary polymer model T, the number of charged AEMA groups remains the same with  polar HEMA groups replacing $\sim 40\%$ of the hydrophobic EMA groups as compared to binary model. With the reduction in number of EMA groups in the ternary polymers, it is important to probe the aggregation dynamics and the effective attractive interactions in the aggregate. The HEMA groups contribute to both electrostatic and van der Waals interactions in the aggregate and the results are shown in Fig.\ref{fig:gr}. In Fig. \ref{fig:gr}(a), the radial distribution function, $g(r)$, between HEMA-AEMA and HEMA-EMA groups strongly exhibits the robust attractive electrostatic interactions between the hydroxyl groups of HEMA and amide groups of AEMA in the aggregate. This underscores the role of effective attractive electrostatic interactions between the model T polymers in offsetting the repulsive interactions between the AEMA groups. This effect can be captured in the computation of electrostatic interactions between AEMA and HEMA groups, as shown in Fig. \ref{fig:gr}(b) plotted as interaction energy, which shows that the effective electrostatic interactions between HEMA-HEMA groups and HEMA-AEMA groups are attractive in nature. The repulsive electrostatic interactions between AEMA groups are also shown for reference. 

Regarding the van der Waals interactions between EMA groups in model B and between (EMA+HEMA) groups in model T, Fig.~\ref{fig:gr}(c) shows that even after replacing $40\%$ of EMA groups with HEMA groups, the van der Waals interactions per non-charged functional groups are not affected significantly. As can be seen from Figure ~\ref{fig:chem}, the polar HEMA groups have the same side group as hydrophobic EMA groups, albeit an addition of OH moiety, suggesting that HEMA groups also contribute to the overall hydrophobic content of the polymer. So, while there is significant difference in EMA-EMA group interaction energies in models B and T (as seen in Fig. \ref{fig:energy1}(a)), if we compute the effective hydrophobic interaction energy by including HEMA groups as well (Fig.~\ref{fig:gr}(c)), the difference between models B and T is significantly reduced. Given that the polymers considered here have $30\%$ charged functional groups, the interactions that drive the aggregation of the polymers in solution have to be van der Waals interactions, which should be able to efficiently overcome the repulsive electrostatic interactions. However, for the polymers like AMpolymers, whose functional goal is to disrupt the bacterial membranes as well as be non-toxic to mammalian membranes, a subtle balance of hydrophobic content with effective weak intra-aggregate interaction is needed for efficient action as membrane active agents. In this regard, the analysis presented here strongly supports the role of inclusion of polar HEMA groups as a tuning parameter for aggregate formation as well as for formation of weaker aggregates. By contributing to the overall hydrophobicity, though smeared along the polymer, the HEMA groups contribute effectively towards formation of aggregates, overcoming the charged group repulsion. They also form attractive electrostatic interactions with the charged groups, effectively neutralising them and contribute further to aggregate stability. However, due to their higher propensity to water and lowered individual hydrophobic content, the HEMA groups preclude the polymers from forming a strong hydrophobic core as can be seen in the binary model B polymers.  The present simulations clearly depict that replacing the strongly hydrophobic  EMA groups by polar HEMA groups helps in 'smearing' the overall hydrophobicity while retaining the overall non-bonded energy and rendering the aggregates of such polymers optimally weaker in solvent than those with a higher content of hydrophobic groups. This should be reflected in how the aggregates of model T and model B polymers effectively interact with bacterial membranes and consequent partitioning of the polymers from the aggregate into the membrane.

\section{\label{sec:discuss}Discussion and Conclusion}
In this study we have investigated the role played by inclusion of polar groups into methacrylate copolymers on the aggregation dynamics in solution phase using atomistic molecular dynamics simulation. The four model polymers considered in this study have a flexible backbone and consist of charged and hydrophobic (binary polymers) or charged, polar and hydrophobic subunits (ternary polymers). The components of the copolymers are either placed randomly (models B and T) or in block sequence of groups (models BB and TB) along the polymer backbone. The chosen composition of ternary copolymers has been shown experimentally to be optimal for antimicrobial activity and also shows significantly reduced hemolytic activity \cite{Mortazavian18} and one of the main goals of the present study is to understand the role played by inclusion of polar HEMA groups on the pre-membrane interaction stage which is the multiple ternary polymers interactions in solution phase.  Both binary and ternary polymers form aggregates in water while displaying markedly different aggregation dynamics. We analyse the morphology of the aggregates and discuss how the presence of polar groups and the role of sequence of various subunits along the polymer backbone affect the aggregation conformation in solution. Our data suggests that block polymers tend to form larger sized aggregates compared to their random counterparts. While the aggregates of only charged and hydrophobic functional groups form well packed, compact aggregates, there are differences with respect to the sequence of such groups along the polymer backbone: block binary (model BB) aggregate displays a spherical conformation with a remarkably strong hydrophobic core which is almost impermeable to water surrounded by charged groups, the random binary (model B) aggregate displays a more ellipsoidal shape with relatively weaker core displaying higher permeability to water. This observation is in very good agreement with scattering experiments and fluorescence imaging of aqueous block and random copolymers, which suggest that intermolecular critical aggregation concentration (CAC) of the random copolymers is much higher than that for the aqueous block copolymers, displaying weaker hydrophobicity in the random copolymer model \cite{Oda11}. We note that the dense hydrophobic core, surrounded by charged units, can help in shielding the hydrophobicity of block copolymers and reduce their non-specific hydrophobic binding to the mammalian membranes, preventing undesirable hemolytic activity. On the other hand, experimental studies have exhibited that random copolymers might not be able to effectively hide the hydrophobicity of copolymer, as indeed can be deduced from our analysis, and may thus bind to human membrane and cause hemolysis \cite{Oda11,Oda18}. 

We then exhibited that the presence of polar HEMA groups induces conformational fluctuations and in general, results in the formation of loosely packed, somewhat extended aggregates, particularly in the case when the groups are randomly distributed along the polymer backbone. Analysing the effect of polar subunits on the binding energies of the inter-polymers in the aggregate showed that replacing the strongly hydrophobic groups by polar groups disperses the overall hydrophobicity of the aggregates while retaining the total attractive van der Waals interactions and counteracting the strong repulsive electrostatic interactions between the charged groups. The HEMA groups also form attractive electrostatic interactions with both charged AEMA and polar HEMA groups further contributing to aggregate formation. Nonetheless, the overall hydrophobicity which drives the aggregation binding and stability is significantly reduced in the ternary aggregates. Such polymers therefore make weaker aggregates in solution phase than those with a higher content of hydrophobic groups. This can have a profound effect on the interaction of the polymers with bacterial membranes as it has been shown that one of the driving forces for partitioning of the polymers from aggregates into the membrane is the relative interaction energy experienced by the polymer in the aggregate and the polymer-membrane interaction energy \cite{baul14,Horn12}. The weak ternary aggregates, upon interacting with bacterial cell membrane, might therefore drive faster dissociation of polymers from the aggregate and consequent polymer insertion and membrane disruption, due to polymer-lipid head groups interactions becoming favourable over polymer-polymer interaction more easily, compared to the densely packed, stable binary aggregates. Therefore, such weak aggregates can positively impact the lysing action of synthetic antibacterial copolymers on bacterial cell membranes.

The results presented in this study delineates the possible role of the neutral polar groups as a modulator which can suitably tune the effect of hydrophobic and charged groups in the polymer aggregates, by smearing the overall hydrophobicity in the aggregate and compensating for the strong repulsive interactions between the charged groups. Our data shows that, under fixed charged subunits composition (comparable to net positive charge of natural AMP\cite{zasloff02}) and same degree of polymerization, ternary copolymer systems can be considered as better biomimetic antibacterial copolymers compared to binary systems with only charged and hydrophobic moieties. This is because the compact binary random assemblies with higher proportion of hydrophobic constituents  might cause undesired toxicity to humans due to their non-specific hydrophobic interactions. On the other hand, while the hemolytic activity of block binary copolymers is reduced due to more effective confinement of the hydrophobic groups, the aggregates formed in this case are densely packed and spherical shaped with strong hydrophobic cores. Such well packed, stable aggregates can inhibit the dissociation of polymers from the aggregate, negatively impacting their antibacterial action, which results from membrane disruption due to insertion of dissociated polymers. This is in broad agreement with experimental studies \cite{Kuroda09,Palermo12} which have observed the levelling off of antimicrobial activity for highly hydrophobic polymers. In our previous work \cite{baul14,baul17}, utilizing MD simulations, we had studied the interaction of aggregates of E4 polymers having charged and hydrophobic groups as constitutents, with the bacterial membrane patch. It was shown in this case that polymers are released into the bilayer from the aggregate due to weak polymer-polymer interactions, which are overcome by polymer-anionic lipid interactions. However, E4 polymer have 7 charged ammonium charged groups and 3 hydrophobic groups in a polymer chain (with degree of polymerization = 10). The smaller hydrophobic content of the studied E4 polymers in the previous studies, mentioned above, can also explain the weaker aggregate formation resulting in efficient membrane partitioning.  However, most natural AMPs \cite{zasloff02} have only $30\%$ net positive charge and experimental studies have also indicated that anti-microbial activity of polymers starts to level off when charged groups are more than $\sim 30\%$ of composition \cite{Mortazavian18}. Therefore, some charged groups in E4 polymer may be excessive and not necessarily required for potent anti microbial activity. On the other hand, a high hydrophobic content can lead to strong aggregate formation decreasing the efficacy of the AM polymers. To balance these two aspects our present data elucidates the critical role played by the introduction of neutral hydrophilic groups into synthetic AM polymers leading to effective aggregation dynamics, consequently on the antibacterial action of such polymers. In particular, our work conclusively highlights the importance of optimum proportion of charged and hydrophobic groups, in addition to the specific role of polar HEMA groups, in reducing the formation of inter polymer contacts of strong hydrophobic groups, thus affecting the aggregation dynamics of ternary copolymers in solution. The results also show that apart from the composition of the AM polymers, the sequence of the functional groups also plays an important role in not only the aggregate formation but in the intra-aggregate interactions resulting in either weak or strong aggregates. This can result in an enhancement of the antibacterial activity and reduction of hemolytic activity of such biomimetic polymers in membrane environment. We also note that natural AMPs have several more functional residues other than only charged, polar, and hydrophobic units \cite{Sato06,Pistolesi07,Colak09,Alvarez15,Wong16}. The design of polymers, mimicking AMPs in displaying optimised compositions of a multitude of subunits, is of particular significance since it can lead to the development of potent antibacterial agents that can act against a broad spectrum of bacteria. Our present work paves the way for the design of such polymers.

\begin{acknowledgments}
All the simulations in this work have been carried out on clusters Annapurna and Nandadevi at The Institute of Mathematical Sciences, Chennai, India.
\end{acknowledgments}

\section*{Data Availability}
The data that support the findings of this study are available from the corresponding author upon reasonable request.
\nocite{*}
\bibliography{AM}

\begin{thebibliography}{58}%
\makeatletter
\providecommand \@ifxundefined [1]{%
 \@ifx{#1\undefined}
}%
\providecommand \@ifnum [1]{%
 \ifnum #1\expandafter \@firstoftwo
 \else \expandafter \@secondoftwo
 \fi
}%
\providecommand \@ifx [1]{%
 \ifx #1\expandafter \@firstoftwo
 \else \expandafter \@secondoftwo
 \fi
}%
\providecommand \natexlab [1]{#1}%
\providecommand \enquote  [1]{``#1''}%
\providecommand \bibnamefont  [1]{#1}%
\providecommand \bibfnamefont [1]{#1}%
\providecommand \citenamefont [1]{#1}%
\providecommand \href@noop [0]{\@secondoftwo}%
\providecommand \href [0]{\begingroup \@sanitize@url \@href}%
\providecommand \@href[1]{\@@startlink{#1}\@@href}%
\providecommand \@@href[1]{\endgroup#1\@@endlink}%
\providecommand \@sanitize@url [0]{\catcode `\\12\catcode `\$12\catcode
  `\&12\catcode `\#12\catcode `\^12\catcode `\_12\catcode `\%12\relax}%
\providecommand \@@startlink[1]{}%
\providecommand \@@endlink[0]{}%
\providecommand \url  [0]{\begingroup\@sanitize@url \@url }%
\providecommand \@url [1]{\endgroup\@href {#1}{\urlprefix }}%
\providecommand \urlprefix  [0]{URL }%
\providecommand \Eprint [0]{\href }%
\providecommand \doibase [0]{http://dx.doi.org/}%
\providecommand \selectlanguage [0]{\@gobble}%
\providecommand \bibinfo  [0]{\@secondoftwo}%
\providecommand \bibfield  [0]{\@secondoftwo}%
\providecommand \translation [1]{[#1]}%
\providecommand \BibitemOpen [0]{}%
\providecommand \bibitemStop [0]{}%
\providecommand \bibitemNoStop [0]{.\EOS\space}%
\providecommand \EOS [0]{\spacefactor3000\relax}%
\providecommand \BibitemShut  [1]{\csname bibitem#1\endcsname}%
\let\auto@bib@innerbib\@empty
\bibitem [{\citenamefont {Locock}\ \emph {et~al.}(2014)\citenamefont {Locock},
  \citenamefont {Michl}, \citenamefont {Stevens}, \citenamefont {Hayball},
  \citenamefont {Vasilev}, \citenamefont {Postma}, \citenamefont {Griesser},
  \citenamefont {Meagher},\ and\ \citenamefont {Haeussler}}]{Locock14}%
  \BibitemOpen
  \bibfield  {author} {\bibinfo {author} {\bibfnamefont {K.~E.~S.}\
  \bibnamefont {Locock}}, \bibinfo {author} {\bibfnamefont {T.~D.}\
  \bibnamefont {Michl}}, \bibinfo {author} {\bibfnamefont {N.}~\bibnamefont
  {Stevens}}, \bibinfo {author} {\bibfnamefont {J.~D.}\ \bibnamefont
  {Hayball}}, \bibinfo {author} {\bibfnamefont {K.}~\bibnamefont {Vasilev}},
  \bibinfo {author} {\bibfnamefont {A.}~\bibnamefont {Postma}}, \bibinfo
  {author} {\bibfnamefont {H.~J.}\ \bibnamefont {Griesser}}, \bibinfo {author}
  {\bibfnamefont {L.}~\bibnamefont {Meagher}}, \ and\ \bibinfo {author}
  {\bibfnamefont {M.}~\bibnamefont {Haeussler}},\ }\bibfield  {title} {\enquote
  {\bibinfo {title} {Antimicrobial polymethacrylates synthesized as mimics of
  tryptophan-rich cationic peptides},}\ }\href {\doibase 10.1021/mz5001527}
  {\bibfield  {journal} {\bibinfo  {journal} {ACS Macro Letters}\ }\textbf
  {\bibinfo {volume} {3}},\ \bibinfo {pages} {319--323} (\bibinfo {year}
  {2014})}\BibitemShut {NoStop}%
\bibitem [{\citenamefont {Takahashi}\ \emph {et~al.}(2017)\citenamefont
  {Takahashi}, \citenamefont {Caputo}, \citenamefont {Vemparala},\ and\
  \citenamefont {Kuroda}}]{Takahashi17}%
  \BibitemOpen
  \bibfield  {author} {\bibinfo {author} {\bibfnamefont {H.}~\bibnamefont
  {Takahashi}}, \bibinfo {author} {\bibfnamefont {G.~A.}\ \bibnamefont
  {Caputo}}, \bibinfo {author} {\bibfnamefont {S.}~\bibnamefont {Vemparala}}, \
  and\ \bibinfo {author} {\bibfnamefont {K.}~\bibnamefont {Kuroda}},\
  }\bibfield  {title} {\enquote {\bibinfo {title} {Synthetic random copolymers
  as a molecular platform to mimic host-defense antimicrobial peptides},}\
  }\href {\doibase 10.1021/acs.bioconjchem.7b00114} {\bibfield  {journal}
  {\bibinfo  {journal} {Bioconjugate Chemistry}\ }\textbf {\bibinfo {volume}
  {28}},\ \bibinfo {pages} {1340--1350} (\bibinfo {year} {2017})}\BibitemShut
  {NoStop}%
\bibitem [{\citenamefont {Ageitos}\ \emph {et~al.}(2017)\citenamefont
  {Ageitos}, \citenamefont {S{\'a}nchez-P{\'e}rez}, \citenamefont {Calo-Mata},\
  and\ \citenamefont {Villa}}]{ageitos2017antimicrobial}%
  \BibitemOpen
  \bibfield  {author} {\bibinfo {author} {\bibfnamefont {J.}~\bibnamefont
  {Ageitos}}, \bibinfo {author} {\bibfnamefont {A.}~\bibnamefont
  {S{\'a}nchez-P{\'e}rez}}, \bibinfo {author} {\bibfnamefont {P.}~\bibnamefont
  {Calo-Mata}}, \ and\ \bibinfo {author} {\bibfnamefont {T.}~\bibnamefont
  {Villa}},\ }\bibfield  {title} {\enquote {\bibinfo {title} {Antimicrobial
  peptides (amps): ancient compounds that represent novel weapons in the fight
  against bacteria},}\ }\href@noop {} {\bibfield  {journal} {\bibinfo
  {journal} {Biochemical pharmacology}\ }\textbf {\bibinfo {volume} {133}},\
  \bibinfo {pages} {117--138} (\bibinfo {year} {2017})}\BibitemShut {NoStop}%
\bibitem [{\citenamefont {Mcphee}\ and\ \citenamefont
  {Hancock}(2005)}]{REVmcphee05}%
  \BibitemOpen
  \bibfield  {author} {\bibinfo {author} {\bibfnamefont {J.~B.}\ \bibnamefont
  {Mcphee}}\ and\ \bibinfo {author} {\bibfnamefont {R.~E.~W.}\ \bibnamefont
  {Hancock}},\ }\bibfield  {title} {\enquote {\bibinfo {title} {Function and
  therapeutic potential of host defence peptides},}\ }\href {\doibase
  10.1002/psc.704} {\bibfield  {journal} {\bibinfo  {journal} {Journal of
  Peptide Science}\ }\textbf {\bibinfo {volume} {11}},\ \bibinfo {pages}
  {677--687} (\bibinfo {year} {2005})}\BibitemShut {NoStop}%
\bibitem [{\citenamefont {Boman}(1995)}]{REVboman95}%
  \BibitemOpen
  \bibfield  {author} {\bibinfo {author} {\bibfnamefont {H.~G.}\ \bibnamefont
  {Boman}},\ }\bibfield  {title} {\enquote {\bibinfo {title} {Peptide
  antibiotics and their role in innate immunity},}\ }\href
  {http://www.ncbi.nlm.nih.gov/pubmed/7612236} {\bibfield  {journal} {\bibinfo
  {journal} {Annu Rev Immunol}\ }\textbf {\bibinfo {volume} {13}},\ \bibinfo
  {pages} {61--92} (\bibinfo {year} {1995})}\BibitemShut {NoStop}%
\bibitem [{\citenamefont {Hwang}\ and\ \citenamefont
  {Vogel}(1998)}]{hwang1998structure}%
  \BibitemOpen
  \bibfield  {author} {\bibinfo {author} {\bibfnamefont {P.~M.}\ \bibnamefont
  {Hwang}}\ and\ \bibinfo {author} {\bibfnamefont {H.~J.}\ \bibnamefont
  {Vogel}},\ }\bibfield  {title} {\enquote {\bibinfo {title}
  {Structure-function relationships of antimicrobial peptides},}\ }\href@noop
  {} {\bibfield  {journal} {\bibinfo  {journal} {Biochemistry and Cell
  Biology}\ }\textbf {\bibinfo {volume} {76}},\ \bibinfo {pages} {235--246}
  (\bibinfo {year} {1998})}\BibitemShut {NoStop}%
\bibitem [{\citenamefont {Zhang}\ and\ \citenamefont
  {Gallo}(2016)}]{zhang2016antimicrobial}%
  \BibitemOpen
  \bibfield  {author} {\bibinfo {author} {\bibfnamefont {L.-j.}\ \bibnamefont
  {Zhang}}\ and\ \bibinfo {author} {\bibfnamefont {R.~L.}\ \bibnamefont
  {Gallo}},\ }\bibfield  {title} {\enquote {\bibinfo {title} {Antimicrobial
  peptides},}\ }\href@noop {} {\bibfield  {journal} {\bibinfo  {journal}
  {Current Biology}\ }\textbf {\bibinfo {volume} {26}},\ \bibinfo {pages}
  {R14--R19} (\bibinfo {year} {2016})}\BibitemShut {NoStop}%
\bibitem [{\citenamefont {Pasupuleti}, \citenamefont {Schmidtchen},\ and\
  \citenamefont {Malmsten}(2012)}]{pasupuleti2012antimicrobial}%
  \BibitemOpen
  \bibfield  {author} {\bibinfo {author} {\bibfnamefont {M.}~\bibnamefont
  {Pasupuleti}}, \bibinfo {author} {\bibfnamefont {A.}~\bibnamefont
  {Schmidtchen}}, \ and\ \bibinfo {author} {\bibfnamefont {M.}~\bibnamefont
  {Malmsten}},\ }\bibfield  {title} {\enquote {\bibinfo {title} {Antimicrobial
  peptides: key components of the innate immune system},}\ }\href@noop {}
  {\bibfield  {journal} {\bibinfo  {journal} {Critical reviews in
  biotechnology}\ }\textbf {\bibinfo {volume} {32}},\ \bibinfo {pages}
  {143--171} (\bibinfo {year} {2012})}\BibitemShut {NoStop}%
\bibitem [{\citenamefont {Nguyen}, \citenamefont {Haney},\ and\ \citenamefont
  {Vogel}(2011)}]{REVnguyen11}%
  \BibitemOpen
  \bibfield  {author} {\bibinfo {author} {\bibfnamefont {L.~T.}\ \bibnamefont
  {Nguyen}}, \bibinfo {author} {\bibfnamefont {E.~F.}\ \bibnamefont {Haney}}, \
  and\ \bibinfo {author} {\bibfnamefont {H.~J.}\ \bibnamefont {Vogel}},\
  }\bibfield  {title} {\enquote {\bibinfo {title} {The expanding scope of
  antimicrobial peptide structures and their modes of action},}\ }\href
  {\doibase http://dx.doi.org/10.1016/j.tibtech.2011.05.001} {\bibfield
  {journal} {\bibinfo  {journal} {Trends in Biotechnology}\ }\textbf {\bibinfo
  {volume} {29}},\ \bibinfo {pages} {464--472} (\bibinfo {year}
  {2011})}\BibitemShut {NoStop}%
\bibitem [{\citenamefont {Tang}\ \emph {et~al.}(2006)\citenamefont {Tang},
  \citenamefont {Doerksen}, \citenamefont {Jones}, \citenamefont {Klein},\ and\
  \citenamefont {Tew}}]{Tang06}%
  \BibitemOpen
  \bibfield  {author} {\bibinfo {author} {\bibfnamefont {H.}~\bibnamefont
  {Tang}}, \bibinfo {author} {\bibfnamefont {R.~J.}\ \bibnamefont {Doerksen}},
  \bibinfo {author} {\bibfnamefont {T.~V.}\ \bibnamefont {Jones}}, \bibinfo
  {author} {\bibfnamefont {M.~L.}\ \bibnamefont {Klein}}, \ and\ \bibinfo
  {author} {\bibfnamefont {G.~N.}\ \bibnamefont {Tew}},\ }\bibfield  {title}
  {\enquote {\bibinfo {title} {Biomimetic facially amphiphilic antibacterial
  oligomers with conformationally stiff backbones},}\ }\href {\doibase
  10.1016/j.chembiol.2006.02.007} {\bibfield  {journal} {\bibinfo  {journal}
  {Chemistry \& Biology}\ }\textbf {\bibinfo {volume} {13}},\ \bibinfo {pages}
  {427 -- 435} (\bibinfo {year} {2006})}\BibitemShut {NoStop}%
\bibitem [{\citenamefont {Som}\ \emph {et~al.}(2008)\citenamefont {Som},
  \citenamefont {Vemparala}, \citenamefont {Ivanov},\ and\ \citenamefont
  {Tew}}]{Som2008}%
  \BibitemOpen
  \bibfield  {author} {\bibinfo {author} {\bibfnamefont {A.}~\bibnamefont
  {Som}}, \bibinfo {author} {\bibfnamefont {S.}~\bibnamefont {Vemparala}},
  \bibinfo {author} {\bibfnamefont {I.}~\bibnamefont {Ivanov}}, \ and\ \bibinfo
  {author} {\bibfnamefont {G.~N.}\ \bibnamefont {Tew}},\ }\bibfield  {title}
  {\enquote {\bibinfo {title} {Synthetic mimics of antimicrobial peptides},}\
  }\href {\doibase 10.1002/bip.20970} {\bibfield  {journal} {\bibinfo
  {journal} {Peptide Science}\ }\textbf {\bibinfo {volume} {90}},\ \bibinfo
  {pages} {83--93} (\bibinfo {year} {2008})}\BibitemShut {NoStop}%
\bibitem [{\citenamefont {Tew}\ \emph {et~al.}(2002)\citenamefont {Tew},
  \citenamefont {Liu}, \citenamefont {Chen}, \citenamefont {Doerksen},
  \citenamefont {Kaplan}, \citenamefont {Carroll}, \citenamefont {Klein},\ and\
  \citenamefont {DeGrado}}]{Tew2002}%
  \BibitemOpen
  \bibfield  {author} {\bibinfo {author} {\bibfnamefont {G.~N.}\ \bibnamefont
  {Tew}}, \bibinfo {author} {\bibfnamefont {D.}~\bibnamefont {Liu}}, \bibinfo
  {author} {\bibfnamefont {B.}~\bibnamefont {Chen}}, \bibinfo {author}
  {\bibfnamefont {R.~J.}\ \bibnamefont {Doerksen}}, \bibinfo {author}
  {\bibfnamefont {J.}~\bibnamefont {Kaplan}}, \bibinfo {author} {\bibfnamefont
  {P.~J.}\ \bibnamefont {Carroll}}, \bibinfo {author} {\bibfnamefont {M.~L.}\
  \bibnamefont {Klein}}, \ and\ \bibinfo {author} {\bibfnamefont {W.~F.}\
  \bibnamefont {DeGrado}},\ }\bibfield  {title} {\enquote {\bibinfo {title} {De
  novo design of biomimetic antimicrobial polymers},}\ }\href {\doibase
  10.1073/pnas.082046199} {\bibfield  {journal} {\bibinfo  {journal}
  {Proceedings of the National Academy of Sciences}\ }\textbf {\bibinfo
  {volume} {99}},\ \bibinfo {pages} {5110--5114} (\bibinfo {year}
  {2002})}\BibitemShut {NoStop}%
\bibitem [{\citenamefont {Ergene}, \citenamefont {Yasuhara},\ and\
  \citenamefont {Palermo}(2018)}]{Palermo18}%
  \BibitemOpen
  \bibfield  {author} {\bibinfo {author} {\bibfnamefont {C.}~\bibnamefont
  {Ergene}}, \bibinfo {author} {\bibfnamefont {K.}~\bibnamefont {Yasuhara}}, \
  and\ \bibinfo {author} {\bibfnamefont {E.~F.}\ \bibnamefont {Palermo}},\
  }\bibfield  {title} {\enquote {\bibinfo {title} {Biomimetic antimicrobial
  polymers: recent advances in molecular design},}\ }\href {\doibase
  10.1039/C8PY00012C} {\bibfield  {journal} {\bibinfo  {journal} {Polym.
  Chem.}\ }\textbf {\bibinfo {volume} {9}},\ \bibinfo {pages} {2407--2427}
  (\bibinfo {year} {2018})}\BibitemShut {NoStop}%
\bibitem [{\citenamefont {Zasloff}(2002)}]{zasloff02}%
  \BibitemOpen
  \bibfield  {author} {\bibinfo {author} {\bibfnamefont {M.}~\bibnamefont
  {Zasloff}},\ }\bibfield  {title} {\enquote {\bibinfo {title} {Antimicrobial
  peptides of multicellular organisms},}\ }\href {\doibase
  http://dx.doi.org/10.1038/415389a} {\bibfield  {journal} {\bibinfo  {journal}
  {Nature}\ }\textbf {\bibinfo {volume} {415}},\ \bibinfo {pages} {389--395}
  (\bibinfo {year} {2002})}\BibitemShut {NoStop}%
\bibitem [{\citenamefont {Liu}\ \emph {et~al.}(2004)\citenamefont {Liu},
  \citenamefont {Choi}, \citenamefont {Chen}, \citenamefont {Doerksen},
  \citenamefont {Clements}, \citenamefont {Winkler}, \citenamefont {Klein},\
  and\ \citenamefont {DeGrado}}]{aryl2}%
  \BibitemOpen
  \bibfield  {author} {\bibinfo {author} {\bibfnamefont {D.}~\bibnamefont
  {Liu}}, \bibinfo {author} {\bibfnamefont {S.}~\bibnamefont {Choi}}, \bibinfo
  {author} {\bibfnamefont {B.}~\bibnamefont {Chen}}, \bibinfo {author}
  {\bibfnamefont {R.~J.}\ \bibnamefont {Doerksen}}, \bibinfo {author}
  {\bibfnamefont {D.~J.}\ \bibnamefont {Clements}}, \bibinfo {author}
  {\bibfnamefont {J.~D.}\ \bibnamefont {Winkler}}, \bibinfo {author}
  {\bibfnamefont {M.~L.}\ \bibnamefont {Klein}}, \ and\ \bibinfo {author}
  {\bibfnamefont {W.~F.}\ \bibnamefont {DeGrado}},\ }\bibfield  {title}
  {\enquote {\bibinfo {title} {Nontoxic membrane-active antimicrobial arylamide
  oligomers},}\ }\href {\doibase 10.1002/anie.200352791} {\bibfield  {journal}
  {\bibinfo  {journal} {Angew. Chem., Int. Ed.}\ }\textbf {\bibinfo {volume}
  {43}},\ \bibinfo {pages} {1158--1162} (\bibinfo {year} {2004})}\BibitemShut
  {NoStop}%
\bibitem [{\citenamefont {Ilker}\ \emph {et~al.}(2004)\citenamefont {Ilker},
  \citenamefont {Nüsslein}, \citenamefont {Tew},\ and\ \citenamefont
  {Coughlin}}]{Firat04}%
  \BibitemOpen
  \bibfield  {author} {\bibinfo {author} {\bibfnamefont {M.~F.}\ \bibnamefont
  {Ilker}}, \bibinfo {author} {\bibfnamefont {K.}~\bibnamefont {Nüsslein}},
  \bibinfo {author} {\bibfnamefont {G.~N.}\ \bibnamefont {Tew}}, \ and\
  \bibinfo {author} {\bibfnamefont {E.~B.}\ \bibnamefont {Coughlin}},\
  }\bibfield  {title} {\enquote {\bibinfo {title} {Tuning the hemolytic and
  antibacterial activities of amphiphilic polynorbornene derivatives},}\ }\href
  {\doibase 10.1021/ja045664d} {\bibfield  {journal} {\bibinfo  {journal}
  {Journal of the American Chemical Society}\ }\textbf {\bibinfo {volume}
  {126}},\ \bibinfo {pages} {15870--15875} (\bibinfo {year}
  {2004})}\BibitemShut {NoStop}%
\bibitem [{\citenamefont {Arnt}, \citenamefont {Nusslein},\ and\ \citenamefont
  {Tew}(2004)}]{phen1}%
  \BibitemOpen
  \bibfield  {author} {\bibinfo {author} {\bibfnamefont {L.}~\bibnamefont
  {Arnt}}, \bibinfo {author} {\bibfnamefont {K.}~\bibnamefont {Nusslein}}, \
  and\ \bibinfo {author} {\bibfnamefont {G.~N.}\ \bibnamefont {Tew}},\
  }\bibfield  {title} {\enquote {\bibinfo {title} {Nonhemolytic abiogenic
  polymers as antimicrobial peptide mimics},}\ }\href {\doibase
  10.1002/pola.20304} {\bibfield  {journal} {\bibinfo  {journal} {J. Polym.
  Sci., Part A: Polym. Chem.}\ }\textbf {\bibinfo {volume} {42}},\ \bibinfo
  {pages} {3860--3864} (\bibinfo {year} {2004})}\BibitemShut {NoStop}%
\bibitem [{\citenamefont {Arnt}\ and\ \citenamefont {Tew}(2002)}]{phen2}%
  \BibitemOpen
  \bibfield  {author} {\bibinfo {author} {\bibfnamefont {L.}~\bibnamefont
  {Arnt}}\ and\ \bibinfo {author} {\bibfnamefont {G.~N.}\ \bibnamefont {Tew}},\
  }\bibfield  {title} {\enquote {\bibinfo {title} {New
  poly(phenyleneethynylene)s with cationic, facially amphiphilic structures},}\
  }\href {\doibase 10.1021/ja026607s} {\bibfield  {journal} {\bibinfo
  {journal} {J. Am. Chem. Soc.}\ }\textbf {\bibinfo {volume} {124}},\ \bibinfo
  {pages} {7664--7665} (\bibinfo {year} {2002})}\BibitemShut {NoStop}%
\bibitem [{\citenamefont {Kuroda}\ and\ \citenamefont
  {DeGrado}(2005)}]{Kenichi05}%
  \BibitemOpen
  \bibfield  {author} {\bibinfo {author} {\bibfnamefont {K.}~\bibnamefont
  {Kuroda}}\ and\ \bibinfo {author} {\bibfnamefont {W.~F.}\ \bibnamefont
  {DeGrado}},\ }\bibfield  {title} {\enquote {\bibinfo {title} {Amphiphilic
  polymethacrylate derivatives as antimicrobial agents},}\ }\href {\doibase
  10.1021/ja044205+} {\bibfield  {journal} {\bibinfo  {journal} {Journal of the
  American Chemical Society}\ }\textbf {\bibinfo {volume} {127}},\ \bibinfo
  {pages} {4128--4129} (\bibinfo {year} {2005})}\BibitemShut {NoStop}%
\bibitem [{\citenamefont {Palermo}, \citenamefont {Vemparala},\ and\
  \citenamefont {Kuroda}(2013)}]{palermobook}%
  \BibitemOpen
  \bibfield  {author} {\bibinfo {author} {\bibfnamefont {E.~F.}\ \bibnamefont
  {Palermo}}, \bibinfo {author} {\bibfnamefont {S.}~\bibnamefont {Vemparala}},
  \ and\ \bibinfo {author} {\bibfnamefont {K.}~\bibnamefont {Kuroda}},\
  }\enquote {\bibinfo {title} {Antimicrobial polymers$:$ molecular design as
  synthetic mimics of host-defense peptides},}\ in\ \href {\doibase
  10.1021/bk-2013-1135.ch019} {\emph {\bibinfo {booktitle} {Tailored Polymer
  Architectures for Pharmaceutical and Biomedical Applications}}}\ (\bibinfo
  {publisher} {American Chemical Society},\ \bibinfo {year} {2013})\
  Chap.~\bibinfo {chapter} {20}, pp.\ \bibinfo {pages} {319--330}\BibitemShut
  {NoStop}%
\bibitem [{\citenamefont {Ivanov}\ \emph {et~al.}(2006)\citenamefont {Ivanov},
  \citenamefont {Vemparala}, \citenamefont {Pophristic}, \citenamefont
  {Kuroda}, \citenamefont {DeGrado}, \citenamefont {McCammon},\ and\
  \citenamefont {Klein}}]{Ivanov06}%
  \BibitemOpen
  \bibfield  {author} {\bibinfo {author} {\bibfnamefont {I.}~\bibnamefont
  {Ivanov}}, \bibinfo {author} {\bibfnamefont {S.}~\bibnamefont {Vemparala}},
  \bibinfo {author} {\bibfnamefont {V.}~\bibnamefont {Pophristic}}, \bibinfo
  {author} {\bibfnamefont {K.}~\bibnamefont {Kuroda}}, \bibinfo {author}
  {\bibfnamefont {W.~F.}\ \bibnamefont {DeGrado}}, \bibinfo {author}
  {\bibfnamefont {J.~A.}\ \bibnamefont {McCammon}}, \ and\ \bibinfo {author}
  {\bibfnamefont {M.~L.}\ \bibnamefont {Klein}},\ }\bibfield  {title} {\enquote
  {\bibinfo {title} {Characterization of nonbiological antimicrobial polymers
  in aqueous solution and at water-lipid interfaces from all-atom molecular
  dynamics},}\ }\href {\doibase 10.1021/ja0564665} {\bibfield  {journal}
  {\bibinfo  {journal} {Journal of the American Chemical Society}\ }\textbf
  {\bibinfo {volume} {128}},\ \bibinfo {pages} {1778--1779} (\bibinfo {year}
  {2006})}\BibitemShut {NoStop}%
\bibitem [{\citenamefont {Palermo}, \citenamefont {Vemparala},\ and\
  \citenamefont {Kuroda}(2012)}]{Palermo12}%
  \BibitemOpen
  \bibfield  {author} {\bibinfo {author} {\bibfnamefont {E.~F.}\ \bibnamefont
  {Palermo}}, \bibinfo {author} {\bibfnamefont {S.}~\bibnamefont {Vemparala}},
  \ and\ \bibinfo {author} {\bibfnamefont {K.}~\bibnamefont {Kuroda}},\
  }\bibfield  {title} {\enquote {\bibinfo {title} {Cationic spacer arm design
  strategy for control of antimicrobial activity and conformation of
  amphiphilic methacrylate random copolymers},}\ }\href {\doibase
  10.1021/bm300342u} {\bibfield  {journal} {\bibinfo  {journal}
  {Biomacromolecules}\ }\textbf {\bibinfo {volume} {13}},\ \bibinfo {pages}
  {1632--1641} (\bibinfo {year} {2012})}\BibitemShut {NoStop}%
\bibitem [{\citenamefont {Baul}, \citenamefont {Kuroda},\ and\ \citenamefont
  {Vemparala}(2014)}]{baul14}%
  \BibitemOpen
  \bibfield  {author} {\bibinfo {author} {\bibfnamefont {U.}~\bibnamefont
  {Baul}}, \bibinfo {author} {\bibfnamefont {K.}~\bibnamefont {Kuroda}}, \ and\
  \bibinfo {author} {\bibfnamefont {S.}~\bibnamefont {Vemparala}},\ }\bibfield
  {title} {\enquote {\bibinfo {title} {Interaction of multiple biomimetic
  antimicrobial polymers with model bacterial membranes},}\ }\href {\doibase
  10.1063/1.4893440} {\bibfield  {journal} {\bibinfo  {journal} {The Journal of
  Chemical Physics}\ }\textbf {\bibinfo {volume} {141}},\ \bibinfo {pages}
  {084902} (\bibinfo {year} {2014})}\BibitemShut {NoStop}%
\bibitem [{\citenamefont {Kenawy}, \citenamefont {Worley},\ and\ \citenamefont
  {Broughton}(2007)}]{Kenawy07}%
  \BibitemOpen
  \bibfield  {author} {\bibinfo {author} {\bibfnamefont {E.-R.}\ \bibnamefont
  {Kenawy}}, \bibinfo {author} {\bibfnamefont {S.~D.}\ \bibnamefont {Worley}},
  \ and\ \bibinfo {author} {\bibfnamefont {R.}~\bibnamefont {Broughton}},\
  }\bibfield  {title} {\enquote {\bibinfo {title} {The chemistry and
  applications of antimicrobial polymers: A state-of-the-art review},}\ }\href
  {\doibase 10.1021/bm061150q} {\bibfield  {journal} {\bibinfo  {journal}
  {Biomacromolecules}\ }\textbf {\bibinfo {volume} {8}},\ \bibinfo {pages}
  {1359--1384} (\bibinfo {year} {2007})}\BibitemShut {NoStop}%
\bibitem [{\citenamefont {Mowery}\ \emph {et~al.}(2009)\citenamefont {Mowery},
  \citenamefont {Lindner}, \citenamefont {Weisblum}, \citenamefont {Stahl},\
  and\ \citenamefont {Gellman}}]{mowery2009structure}%
  \BibitemOpen
  \bibfield  {author} {\bibinfo {author} {\bibfnamefont {B.~P.}\ \bibnamefont
  {Mowery}}, \bibinfo {author} {\bibfnamefont {A.~H.}\ \bibnamefont {Lindner}},
  \bibinfo {author} {\bibfnamefont {B.}~\bibnamefont {Weisblum}}, \bibinfo
  {author} {\bibfnamefont {S.~S.}\ \bibnamefont {Stahl}}, \ and\ \bibinfo
  {author} {\bibfnamefont {S.~H.}\ \bibnamefont {Gellman}},\ }\bibfield
  {title} {\enquote {\bibinfo {title} {Structure- activity relationships among
  random nylon-3 copolymers that mimic antibacterial host-defense peptides},}\
  }\href@noop {} {\bibfield  {journal} {\bibinfo  {journal} {Journal of the
  American Chemical Society}\ }\textbf {\bibinfo {volume} {131}},\ \bibinfo
  {pages} {9735--9745} (\bibinfo {year} {2009})}\BibitemShut {NoStop}%
\bibitem [{\citenamefont {Oda}\ \emph {et~al.}(2018)\citenamefont {Oda},
  \citenamefont {Yasuhara}, \citenamefont {Kanaoka}, \citenamefont {Sato},
  \citenamefont {Aoshima},\ and\ \citenamefont {Kuroda}}]{Oda18}%
  \BibitemOpen
  \bibfield  {author} {\bibinfo {author} {\bibfnamefont {Y.}~\bibnamefont
  {Oda}}, \bibinfo {author} {\bibfnamefont {K.}~\bibnamefont {Yasuhara}},
  \bibinfo {author} {\bibfnamefont {S.}~\bibnamefont {Kanaoka}}, \bibinfo
  {author} {\bibfnamefont {T.}~\bibnamefont {Sato}}, \bibinfo {author}
  {\bibfnamefont {S.}~\bibnamefont {Aoshima}}, \ and\ \bibinfo {author}
  {\bibfnamefont {K.}~\bibnamefont {Kuroda}},\ }\bibfield  {title} {\enquote
  {\bibinfo {title} {Aggregation of cationic amphiphilic block and random
  copoly(vinyl ether)s with antimicrobial activity},}\ }\href {\doibase
  1093.3390/polym10010093} {\bibfield  {journal} {\bibinfo  {journal}
  {Polymers}\ }\textbf {\bibinfo {volume} {10}},\ \bibinfo {pages} {93}
  (\bibinfo {year} {2018})}\BibitemShut {NoStop}%
\bibitem [{\citenamefont {Sato}\ and\ \citenamefont
  {Matsuda}(2009)}]{sato2009macromolecular}%
  \BibitemOpen
  \bibfield  {author} {\bibinfo {author} {\bibfnamefont {T.}~\bibnamefont
  {Sato}}\ and\ \bibinfo {author} {\bibfnamefont {Y.}~\bibnamefont {Matsuda}},\
  }\bibfield  {title} {\enquote {\bibinfo {title} {Macromolecular assemblies in
  solution: characterization by light scattering},}\ }\href@noop {} {\bibfield
  {journal} {\bibinfo  {journal} {Polymer journal}\ }\textbf {\bibinfo {volume}
  {41}},\ \bibinfo {pages} {241--251} (\bibinfo {year} {2009})}\BibitemShut
  {NoStop}%
\bibitem [{\citenamefont {Nakashima}\ and\ \citenamefont
  {Bahadur}(2006)}]{nakashima2006aggregation}%
  \BibitemOpen
  \bibfield  {author} {\bibinfo {author} {\bibfnamefont {K.}~\bibnamefont
  {Nakashima}}\ and\ \bibinfo {author} {\bibfnamefont {P.}~\bibnamefont
  {Bahadur}},\ }\bibfield  {title} {\enquote {\bibinfo {title} {Aggregation of
  water-soluble block copolymers in aqueous solutions: recent trends},}\
  }\href@noop {} {\bibfield  {journal} {\bibinfo  {journal} {Advances in
  colloid and interface science}\ }\textbf {\bibinfo {volume} {123}},\ \bibinfo
  {pages} {75--96} (\bibinfo {year} {2006})}\BibitemShut {NoStop}%
\bibitem [{\citenamefont {Oda}\ \emph {et~al.}(2011)\citenamefont {Oda},
  \citenamefont {Kanaoka}, \citenamefont {Sato}, \citenamefont {Aoshima},\ and\
  \citenamefont {Kuroda}}]{Oda11}%
  \BibitemOpen
  \bibfield  {author} {\bibinfo {author} {\bibfnamefont {Y.}~\bibnamefont
  {Oda}}, \bibinfo {author} {\bibfnamefont {S.}~\bibnamefont {Kanaoka}},
  \bibinfo {author} {\bibfnamefont {T.}~\bibnamefont {Sato}}, \bibinfo {author}
  {\bibfnamefont {S.}~\bibnamefont {Aoshima}}, \ and\ \bibinfo {author}
  {\bibfnamefont {K.}~\bibnamefont {Kuroda}},\ }\bibfield  {title} {\enquote
  {\bibinfo {title} {Block versus random amphiphilic copolymers as
  antibacterial agents},}\ }\href {\doibase 10.1021/bm200780r} {\bibfield
  {journal} {\bibinfo  {journal} {Biomacromolecules}\ }\textbf {\bibinfo
  {volume} {12}},\ \bibinfo {pages} {3581--3591} (\bibinfo {year}
  {2011})}\BibitemShut {NoStop}%
\bibitem [{\citenamefont {Uppu}\ \emph {et~al.}(2016)\citenamefont {Uppu},
  \citenamefont {Samaddar}, \citenamefont {Hoque}, \citenamefont {Konai},
  \citenamefont {Krishnamoorthy}, \citenamefont {Shome},\ and\ \citenamefont
  {Haldar}}]{Uppu16}%
  \BibitemOpen
  \bibfield  {author} {\bibinfo {author} {\bibfnamefont {D.~S. S.~M.}\
  \bibnamefont {Uppu}}, \bibinfo {author} {\bibfnamefont {S.}~\bibnamefont
  {Samaddar}}, \bibinfo {author} {\bibfnamefont {J.}~\bibnamefont {Hoque}},
  \bibinfo {author} {\bibfnamefont {M.~M.}\ \bibnamefont {Konai}}, \bibinfo
  {author} {\bibfnamefont {P.}~\bibnamefont {Krishnamoorthy}}, \bibinfo
  {author} {\bibfnamefont {B.~R.}\ \bibnamefont {Shome}}, \ and\ \bibinfo
  {author} {\bibfnamefont {J.}~\bibnamefont {Haldar}},\ }\bibfield  {title}
  {\enquote {\bibinfo {title} {Side chain degradable cationic–amphiphilic
  polymers with tunable hydrophobicity show in vivo activity},}\ }\href
  {\doibase 10.1021/acs.biomac.6b01057} {\bibfield  {journal} {\bibinfo
  {journal} {Biomacromolecules}\ }\textbf {\bibinfo {volume} {17}},\ \bibinfo
  {pages} {3094--3102} (\bibinfo {year} {2016})}\BibitemShut {NoStop}%
\bibitem [{\citenamefont {Ganewatta}\ and\ \citenamefont
  {Tang}(2015)}]{Mitra15}%
  \BibitemOpen
  \bibfield  {author} {\bibinfo {author} {\bibfnamefont {M.~S.}\ \bibnamefont
  {Ganewatta}}\ and\ \bibinfo {author} {\bibfnamefont {C.}~\bibnamefont
  {Tang}},\ }\bibfield  {title} {\enquote {\bibinfo {title} {Controlling
  macromolecular structures towards effective antimicrobial polymers},}\ }\href
  {\doibase 10.1016/j.polymer.2015.03.007} {\bibfield  {journal} {\bibinfo
  {journal} {Polymer (Korea)}\ }\textbf {\bibinfo {volume} {63}},\ \bibinfo
  {pages} {A1--A29} (\bibinfo {year} {2015})}\BibitemShut {NoStop}%
\bibitem [{\citenamefont {Yang}\ \emph {et~al.}(2018)\citenamefont {Yang},
  \citenamefont {Cai}, \citenamefont {Huang}, \citenamefont {Tang},\ and\
  \citenamefont {Zhang}}]{Yang18}%
  \BibitemOpen
  \bibfield  {author} {\bibinfo {author} {\bibfnamefont {Y.}~\bibnamefont
  {Yang}}, \bibinfo {author} {\bibfnamefont {Z.}~\bibnamefont {Cai}}, \bibinfo
  {author} {\bibfnamefont {Z.}~\bibnamefont {Huang}}, \bibinfo {author}
  {\bibfnamefont {X.}~\bibnamefont {Tang}}, \ and\ \bibinfo {author}
  {\bibfnamefont {X.}~\bibnamefont {Zhang}},\ }\bibfield  {title} {\enquote
  {\bibinfo {title} {Antimicrobial cationic polymers: from structural design to
  functional control},}\ }\href {\doibase 10.1038/pj.2017.72} {\bibfield
  {journal} {\bibinfo  {journal} {Polymer Journals}\ }\textbf {\bibinfo
  {volume} {50}},\ \bibinfo {pages} {33--44} (\bibinfo {year}
  {2018})}\BibitemShut {NoStop}%
\bibitem [{\citenamefont {Wang}, \citenamefont {Li},\ and\ \citenamefont
  {Wang}(2008)}]{wang08}%
  \BibitemOpen
  \bibfield  {author} {\bibinfo {author} {\bibfnamefont {G.}~\bibnamefont
  {Wang}}, \bibinfo {author} {\bibfnamefont {X.}~\bibnamefont {Li}}, \ and\
  \bibinfo {author} {\bibfnamefont {Z.}~\bibnamefont {Wang}},\ }\bibfield
  {title} {\enquote {\bibinfo {title} {Apd2: the updated antimicrobial peptide
  database and its application in peptide design. nucleic acids res
  37:d933-d937},}\ }\href {\doibase 10.1093/nar/gkn823} {\bibfield  {journal}
  {\bibinfo  {journal} {Nucleic acids research}\ }\textbf {\bibinfo {volume}
  {37}},\ \bibinfo {pages} {D933--7} (\bibinfo {year} {2008})}\BibitemShut
  {NoStop}%
\bibitem [{\citenamefont {Wang}\ and\ \citenamefont
  {Mishra}(2012)}]{wang2012importance}%
  \BibitemOpen
  \bibfield  {author} {\bibinfo {author} {\bibfnamefont {G.}~\bibnamefont
  {Wang}}\ and\ \bibinfo {author} {\bibfnamefont {B.}~\bibnamefont {Mishra}},\
  }\bibfield  {title} {\enquote {\bibinfo {title} {The importance of amino acid
  composition in natural amps: an evolutional, structural, and functional
  perspective},}\ }\href@noop {} {\bibfield  {journal} {\bibinfo  {journal}
  {Frontiers in immunology}\ }\textbf {\bibinfo {volume} {3}},\ \bibinfo
  {pages} {221} (\bibinfo {year} {2012})}\BibitemShut {NoStop}%
\bibitem [{\citenamefont {Chakraborty}\ \emph {et~al.}(2014)\citenamefont
  {Chakraborty}, \citenamefont {Liu}, \citenamefont {Hayouka}, \citenamefont
  {Chen}, \citenamefont {Ehrhardt}, \citenamefont {Lu}, \citenamefont {Burke},
  \citenamefont {Yang}, \citenamefont {Weisblum}, \citenamefont {Wong},
  \citenamefont {Masters},\ and\ \citenamefont {Gellman}}]{Chakraborty14}%
  \BibitemOpen
  \bibfield  {author} {\bibinfo {author} {\bibfnamefont {S.}~\bibnamefont
  {Chakraborty}}, \bibinfo {author} {\bibfnamefont {R.}~\bibnamefont {Liu}},
  \bibinfo {author} {\bibfnamefont {Z.}~\bibnamefont {Hayouka}}, \bibinfo
  {author} {\bibfnamefont {X.}~\bibnamefont {Chen}}, \bibinfo {author}
  {\bibfnamefont {J.}~\bibnamefont {Ehrhardt}}, \bibinfo {author}
  {\bibfnamefont {Q.}~\bibnamefont {Lu}}, \bibinfo {author} {\bibfnamefont
  {E.}~\bibnamefont {Burke}}, \bibinfo {author} {\bibfnamefont
  {Y.}~\bibnamefont {Yang}}, \bibinfo {author} {\bibfnamefont {B.}~\bibnamefont
  {Weisblum}}, \bibinfo {author} {\bibfnamefont {G.~C.~L.}\ \bibnamefont
  {Wong}}, \bibinfo {author} {\bibfnamefont {K.~S.}\ \bibnamefont {Masters}}, \
  and\ \bibinfo {author} {\bibfnamefont {S.~H.}\ \bibnamefont {Gellman}},\
  }\bibfield  {title} {\enquote {\bibinfo {title} {Ternary nylon-3 copolymers
  as host-defense peptide mimics: Beyond hydrophobic and cationic subunits},}\
  }\href {\doibase 10.1021/ja507576a} {\bibfield  {journal} {\bibinfo
  {journal} {Journal of the American Chemical Society}\ }\textbf {\bibinfo
  {volume} {136}},\ \bibinfo {pages} {14530--14535} (\bibinfo {year}
  {2014})}\BibitemShut {NoStop}%
\bibitem [{\citenamefont {Yang}\ \emph {et~al.}(2014)\citenamefont {Yang},
  \citenamefont {Hu}, \citenamefont {Hu}, \citenamefont {Shi}, \citenamefont
  {Jiang}, \citenamefont {Hui}, \citenamefont {Zhu}, \citenamefont {Xie},\ and\
  \citenamefont {Yang}}]{Yang14}%
  \BibitemOpen
  \bibfield  {author} {\bibinfo {author} {\bibfnamefont {X.}~\bibnamefont
  {Yang}}, \bibinfo {author} {\bibfnamefont {K.}~\bibnamefont {Hu}}, \bibinfo
  {author} {\bibfnamefont {G.}~\bibnamefont {Hu}}, \bibinfo {author}
  {\bibfnamefont {D.}~\bibnamefont {Shi}}, \bibinfo {author} {\bibfnamefont
  {Y.}~\bibnamefont {Jiang}}, \bibinfo {author} {\bibfnamefont
  {L.}~\bibnamefont {Hui}}, \bibinfo {author} {\bibfnamefont {R.}~\bibnamefont
  {Zhu}}, \bibinfo {author} {\bibfnamefont {Y.}~\bibnamefont {Xie}}, \ and\
  \bibinfo {author} {\bibfnamefont {L.}~\bibnamefont {Yang}},\ }\bibfield
  {title} {\enquote {\bibinfo {title} {Long hydrophilic-and-cationic polymers:
  A different pathway toward preferential activity against bacterial over
  mammalian membranes},}\ }\href {\doibase 10.1021/bm5006596} {\bibfield
  {journal} {\bibinfo  {journal} {Biomacromolecules}\ }\textbf {\bibinfo
  {volume} {15}},\ \bibinfo {pages} {3267--3277} (\bibinfo {year}
  {2014})}\BibitemShut {NoStop}%
\bibitem [{\citenamefont {Gao}\ \emph {et~al.}(2011)\citenamefont {Gao},
  \citenamefont {Lange}, \citenamefont {Hilpert}, \citenamefont {Kindrachuk},
  \citenamefont {Zou}, \citenamefont {Cheng}, \citenamefont
  {Kazemzadeh-Narbat}, \citenamefont {Yu}, \citenamefont {Wang}, \citenamefont
  {Straus}, \citenamefont {Brooks}, \citenamefont {Chew}, \citenamefont
  {Hancock},\ and\ \citenamefont {Kizhakkedathu}}]{Gao11}%
  \BibitemOpen
  \bibfield  {author} {\bibinfo {author} {\bibfnamefont {G.}~\bibnamefont
  {Gao}}, \bibinfo {author} {\bibfnamefont {D.}~\bibnamefont {Lange}}, \bibinfo
  {author} {\bibfnamefont {K.}~\bibnamefont {Hilpert}}, \bibinfo {author}
  {\bibfnamefont {J.}~\bibnamefont {Kindrachuk}}, \bibinfo {author}
  {\bibfnamefont {Y.}~\bibnamefont {Zou}}, \bibinfo {author} {\bibfnamefont
  {J.~T.}\ \bibnamefont {Cheng}}, \bibinfo {author} {\bibfnamefont
  {M.}~\bibnamefont {Kazemzadeh-Narbat}}, \bibinfo {author} {\bibfnamefont
  {K.}~\bibnamefont {Yu}}, \bibinfo {author} {\bibfnamefont {R.}~\bibnamefont
  {Wang}}, \bibinfo {author} {\bibfnamefont {S.~K.}\ \bibnamefont {Straus}},
  \bibinfo {author} {\bibfnamefont {D.~E.}\ \bibnamefont {Brooks}}, \bibinfo
  {author} {\bibfnamefont {B.~H.}\ \bibnamefont {Chew}}, \bibinfo {author}
  {\bibfnamefont {R.~E.}\ \bibnamefont {Hancock}}, \ and\ \bibinfo {author}
  {\bibfnamefont {J.~N.}\ \bibnamefont {Kizhakkedathu}},\ }\bibfield  {title}
  {\enquote {\bibinfo {title} {The biocompatibility and biofilm resistance of
  implant coatings based on hydrophilic polymer brushes conjugated with
  antimicrobial peptides},}\ }\href {\doibase
  https://doi.org/10.1016/j.biomaterials.2011.02.013} {\bibfield  {journal}
  {\bibinfo  {journal} {Biomaterials}\ }\textbf {\bibinfo {volume} {32}},\
  \bibinfo {pages} {3899 -- 3909} (\bibinfo {year} {2011})}\BibitemShut
  {NoStop}%
\bibitem [{\citenamefont {Mortazavian}\ \emph {et~al.}(2018)\citenamefont
  {Mortazavian}, \citenamefont {Foster}, \citenamefont {Bhat}, \citenamefont
  {Patel},\ and\ \citenamefont {Kuroda}}]{Mortazavian18}%
  \BibitemOpen
  \bibfield  {author} {\bibinfo {author} {\bibfnamefont {H.}~\bibnamefont
  {Mortazavian}}, \bibinfo {author} {\bibfnamefont {L.~L.}\ \bibnamefont
  {Foster}}, \bibinfo {author} {\bibfnamefont {R.}~\bibnamefont {Bhat}},
  \bibinfo {author} {\bibfnamefont {S.}~\bibnamefont {Patel}}, \ and\ \bibinfo
  {author} {\bibfnamefont {K.}~\bibnamefont {Kuroda}},\ }\bibfield  {title}
  {\enquote {\bibinfo {title} {Decoupling the functional roles of cationic and
  hydrophobic groups in the antimicrobial and hemolytic activities of
  methacrylate random copolymers},}\ }\href {\doibase
  10.1021/acs.biomac.8b01256} {\bibfield  {journal} {\bibinfo  {journal}
  {Biomacromolecules}\ }\textbf {\bibinfo {volume} {19}},\ \bibinfo {pages}
  {4370--4378} (\bibinfo {year} {2018})}\BibitemShut {NoStop}%
\bibitem [{\citenamefont {Jorgensen}\ \emph {et~al.}(1983)\citenamefont
  {Jorgensen}, \citenamefont {Chandrasekhar}, \citenamefont {Madura},
  \citenamefont {Impey},\ and\ \citenamefont {Klein}}]{jorgTIP83}%
  \BibitemOpen
  \bibfield  {author} {\bibinfo {author} {\bibfnamefont {W.~L.}\ \bibnamefont
  {Jorgensen}}, \bibinfo {author} {\bibfnamefont {J.}~\bibnamefont
  {Chandrasekhar}}, \bibinfo {author} {\bibfnamefont {J.~D.}\ \bibnamefont
  {Madura}}, \bibinfo {author} {\bibfnamefont {R.~W.}\ \bibnamefont {Impey}}, \
  and\ \bibinfo {author} {\bibfnamefont {M.~L.}\ \bibnamefont {Klein}},\
  }\bibfield  {title} {\enquote {\bibinfo {title} {Comparison of simple
  potential functions for simulating liquid water},}\ }\href@noop {} {\bibfield
   {journal} {\bibinfo  {journal} {The Journal of Chemical Physics}\ }\textbf
  {\bibinfo {volume} {79}},\ \bibinfo {pages} {926--935} (\bibinfo {year}
  {1983})}\BibitemShut {NoStop}%
\bibitem [{\citenamefont {Phillips}\ \emph {et~al.}(2005)\citenamefont
  {Phillips}, \citenamefont {Braun}, \citenamefont {Wang}, \citenamefont
  {Gumbart}, \citenamefont {Tajkhorshid}, \citenamefont {Villa}, \citenamefont
  {Chipot}, \citenamefont {Skeel}, \citenamefont {Kale},\ and\ \citenamefont
  {Schulten}}]{Phillips05}%
  \BibitemOpen
  \bibfield  {author} {\bibinfo {author} {\bibfnamefont {J.~C.}\ \bibnamefont
  {Phillips}}, \bibinfo {author} {\bibfnamefont {R.}~\bibnamefont {Braun}},
  \bibinfo {author} {\bibfnamefont {W.}~\bibnamefont {Wang}}, \bibinfo {author}
  {\bibfnamefont {J.}~\bibnamefont {Gumbart}}, \bibinfo {author} {\bibfnamefont
  {E.}~\bibnamefont {Tajkhorshid}}, \bibinfo {author} {\bibfnamefont
  {E.}~\bibnamefont {Villa}}, \bibinfo {author} {\bibfnamefont
  {C.}~\bibnamefont {Chipot}}, \bibinfo {author} {\bibfnamefont {R.~D.}\
  \bibnamefont {Skeel}}, \bibinfo {author} {\bibfnamefont {L.}~\bibnamefont
  {Kale}}, \ and\ \bibinfo {author} {\bibfnamefont {K.}~\bibnamefont
  {Schulten}},\ }\bibfield  {title} {\enquote {\bibinfo {title} {Scalable
  molecular dynamics with {NAMD}},}\ }\href {\doibase 10.1002/jcc.20289}
  {\bibfield  {journal} {\bibinfo  {journal} {Journal of Computational
  Chemistry}\ }\textbf {\bibinfo {volume} {26}},\ \bibinfo {pages} {1781 --
  1802} (\bibinfo {year} {2005})}\BibitemShut {NoStop}%
\bibitem [{\citenamefont {Martyna}, \citenamefont {Tobias},\ and\ \citenamefont
  {Klein}(1994)}]{Martyna94}%
  \BibitemOpen
  \bibfield  {author} {\bibinfo {author} {\bibfnamefont {G.~J.}\ \bibnamefont
  {Martyna}}, \bibinfo {author} {\bibfnamefont {D.~J.}\ \bibnamefont {Tobias}},
  \ and\ \bibinfo {author} {\bibfnamefont {M.~L.}\ \bibnamefont {Klein}},\
  }\bibfield  {title} {\enquote {\bibinfo {title} {Constant pressure molecular
  dynamics algorithms},}\ }\href {\doibase 10.1063/1.467468} {\bibfield
  {journal} {\bibinfo  {journal} {The Journal of Chemical Physics}\ }\textbf
  {\bibinfo {volume} {101}},\ \bibinfo {pages} {4177--4189} (\bibinfo {year}
  {1994})}\BibitemShut {NoStop}%
\bibitem [{\citenamefont {Feller}\ \emph {et~al.}(1995)\citenamefont {Feller},
  \citenamefont {Zhang}, \citenamefont {Pastor},\ and\ \citenamefont
  {Brooks}}]{Feller95}%
  \BibitemOpen
  \bibfield  {author} {\bibinfo {author} {\bibfnamefont {S.~E.}\ \bibnamefont
  {Feller}}, \bibinfo {author} {\bibfnamefont {Y.}~\bibnamefont {Zhang}},
  \bibinfo {author} {\bibfnamefont {R.~W.}\ \bibnamefont {Pastor}}, \ and\
  \bibinfo {author} {\bibfnamefont {B.~R.}\ \bibnamefont {Brooks}},\ }\bibfield
   {title} {\enquote {\bibinfo {title} {Constant pressure molecular dynamics
  simulation$:$ the {L}angevin piston method},}\ }\href@noop {} {\bibfield
  {journal} {\bibinfo  {journal} {The Journal of Chemical Physics}\ }\textbf
  {\bibinfo {volume} {103}},\ \bibinfo {pages} {4613--4621} (\bibinfo {year}
  {1995})}\BibitemShut {NoStop}%
\bibitem [{\citenamefont {Essmann}\ \emph {et~al.}(1995)\citenamefont
  {Essmann}, \citenamefont {Perera}, \citenamefont {Berkowitz}, \citenamefont
  {Darden}, \citenamefont {Lee},\ and\ \citenamefont {Pedersen}}]{Essmann95}%
  \BibitemOpen
  \bibfield  {author} {\bibinfo {author} {\bibfnamefont {U.}~\bibnamefont
  {Essmann}}, \bibinfo {author} {\bibfnamefont {L.}~\bibnamefont {Perera}},
  \bibinfo {author} {\bibfnamefont {M.~L.}\ \bibnamefont {Berkowitz}}, \bibinfo
  {author} {\bibfnamefont {T.}~\bibnamefont {Darden}}, \bibinfo {author}
  {\bibfnamefont {H.}~\bibnamefont {Lee}}, \ and\ \bibinfo {author}
  {\bibfnamefont {L.~G.}\ \bibnamefont {Pedersen}},\ }\bibfield  {title}
  {\enquote {\bibinfo {title} {A smooth particle mesh ewald method},}\ }\href
  {\doibase 10.1063/1.470117} {\bibfield  {journal} {\bibinfo  {journal} {The
  Journal of Chemical Physics}\ }\textbf {\bibinfo {volume} {103}},\ \bibinfo
  {pages} {8577--8593} (\bibinfo {year} {1995})}\BibitemShut {NoStop}%
\bibitem [{\citenamefont {MacKerell}\ \emph {et~al.}(1998)\citenamefont
  {MacKerell}, \citenamefont {Bashford}, \citenamefont {Bellott}, \citenamefont
  {Dunbrack}, \citenamefont {Evanseck}, \citenamefont {Field}, \citenamefont
  {Fischer}, \citenamefont {Gao}, \citenamefont {Guo}, \citenamefont {Ha},
  \citenamefont {Joseph-McCarthy}, \citenamefont {Kuchnir}, \citenamefont
  {Kuczera}, \citenamefont {Lau}, \citenamefont {Mattos}, \citenamefont
  {Michnick}, \citenamefont {Ngo}, \citenamefont {Nguyen}, \citenamefont
  {Prodhom}, \citenamefont {Reiher}, \citenamefont {Roux}, \citenamefont
  {Schlenkrich}, \citenamefont {Smith}, \citenamefont {Stote}, \citenamefont
  {Straub}, \citenamefont {Watanabe}, \citenamefont {Wi{\'o}rkiewicz-Kuczera},
  \citenamefont {Yin},\ and\ \citenamefont {Karplus}}]{karplus98}%
  \BibitemOpen
  \bibfield  {author} {\bibinfo {author} {\bibfnamefont {A.~D.}\ \bibnamefont
  {MacKerell}}, \bibinfo {author} {\bibfnamefont {D.}~\bibnamefont {Bashford}},
  \bibinfo {author} {\bibfnamefont {M.}~\bibnamefont {Bellott}}, \bibinfo
  {author} {\bibfnamefont {R.~L.}\ \bibnamefont {Dunbrack}}, \bibinfo {author}
  {\bibfnamefont {J.~D.}\ \bibnamefont {Evanseck}}, \bibinfo {author}
  {\bibfnamefont {M.~J.}\ \bibnamefont {Field}}, \bibinfo {author}
  {\bibfnamefont {S.}~\bibnamefont {Fischer}}, \bibinfo {author} {\bibfnamefont
  {J.}~\bibnamefont {Gao}}, \bibinfo {author} {\bibfnamefont {H.}~\bibnamefont
  {Guo}}, \bibinfo {author} {\bibfnamefont {S.}~\bibnamefont {Ha}}, \bibinfo
  {author} {\bibfnamefont {D.}~\bibnamefont {Joseph-McCarthy}}, \bibinfo
  {author} {\bibfnamefont {L.}~\bibnamefont {Kuchnir}}, \bibinfo {author}
  {\bibfnamefont {K.}~\bibnamefont {Kuczera}}, \bibinfo {author} {\bibfnamefont
  {F.~T.~K.}\ \bibnamefont {Lau}}, \bibinfo {author} {\bibfnamefont
  {C.}~\bibnamefont {Mattos}}, \bibinfo {author} {\bibfnamefont
  {S.}~\bibnamefont {Michnick}}, \bibinfo {author} {\bibfnamefont
  {T.}~\bibnamefont {Ngo}}, \bibinfo {author} {\bibfnamefont {D.~T.}\
  \bibnamefont {Nguyen}}, \bibinfo {author} {\bibfnamefont {B.}~\bibnamefont
  {Prodhom}}, \bibinfo {author} {\bibfnamefont {W.~E.}\ \bibnamefont {Reiher}},
  \bibinfo {author} {\bibfnamefont {B.}~\bibnamefont {Roux}}, \bibinfo {author}
  {\bibfnamefont {M.}~\bibnamefont {Schlenkrich}}, \bibinfo {author}
  {\bibfnamefont {J.~C.}\ \bibnamefont {Smith}}, \bibinfo {author}
  {\bibfnamefont {R.}~\bibnamefont {Stote}}, \bibinfo {author} {\bibfnamefont
  {J.}~\bibnamefont {Straub}}, \bibinfo {author} {\bibfnamefont
  {M.}~\bibnamefont {Watanabe}}, \bibinfo {author} {\bibfnamefont
  {J.}~\bibnamefont {Wi{\'o}rkiewicz-Kuczera}}, \bibinfo {author}
  {\bibfnamefont {D.}~\bibnamefont {Yin}}, \ and\ \bibinfo {author}
  {\bibfnamefont {M.}~\bibnamefont {Karplus}},\ }\bibfield  {title} {\enquote
  {\bibinfo {title} {All-atom empirical potential for molecular modeling and
  dynamics studies of proteins},}\ }\href {\doibase 10.1021/jp973084f}
  {\bibfield  {journal} {\bibinfo  {journal} {The Journal of Physical Chemistry
  B}\ }\textbf {\bibinfo {volume} {102}},\ \bibinfo {pages} {3586--3616}
  (\bibinfo {year} {1998})}\BibitemShut {NoStop}%
\bibitem [{\citenamefont {Humphrey}, \citenamefont {Dalke},\ and\ \citenamefont
  {Schulten}(1996)}]{vmd96}%
  \BibitemOpen
  \bibfield  {author} {\bibinfo {author} {\bibfnamefont {W.}~\bibnamefont
  {Humphrey}}, \bibinfo {author} {\bibfnamefont {A.}~\bibnamefont {Dalke}}, \
  and\ \bibinfo {author} {\bibfnamefont {K.}~\bibnamefont {Schulten}},\
  }\bibfield  {title} {\enquote {\bibinfo {title} {{VMD}$:$ visual molecular
  dynamics},}\ }\href {\doibase http://dx.doi.org/10.1016/0263-7855(96)00018-5}
  {\bibfield  {journal} {\bibinfo  {journal} {Journal of Molecular Graphics}\
  }\textbf {\bibinfo {volume} {14}},\ \bibinfo {pages} {33 -- 38} (\bibinfo
  {year} {1996})}\BibitemShut {NoStop}%
\bibitem [{\citenamefont {Eismin}\ \emph {et~al.}(2017)\citenamefont {Eismin},
  \citenamefont {Munusamy}, \citenamefont {Kegel}, \citenamefont {Hogan},
  \citenamefont {Maier}, \citenamefont {Schwartz},\ and\ \citenamefont
  {Pemberton}}]{ecc17}%
  \BibitemOpen
  \bibfield  {author} {\bibinfo {author} {\bibfnamefont {R.~J.}\ \bibnamefont
  {Eismin}}, \bibinfo {author} {\bibfnamefont {E.}~\bibnamefont {Munusamy}},
  \bibinfo {author} {\bibfnamefont {L.~L.}\ \bibnamefont {Kegel}}, \bibinfo
  {author} {\bibfnamefont {D.~E.}\ \bibnamefont {Hogan}}, \bibinfo {author}
  {\bibfnamefont {R.~M.}\ \bibnamefont {Maier}}, \bibinfo {author}
  {\bibfnamefont {S.~D.}\ \bibnamefont {Schwartz}}, \ and\ \bibinfo {author}
  {\bibfnamefont {J.~E.}\ \bibnamefont {Pemberton}},\ }\bibfield  {title}
  {\enquote {\bibinfo {title} {Evolution of aggregate structure in solutions of
  anionic monorhamnolipids: Experimental and computational results},}\ }\href
  {\doibase 10.1021/acs.langmuir.7b00078} {\bibfield  {journal} {\bibinfo
  {journal} {Langmuir}\ }\textbf {\bibinfo {volume} {33}},\ \bibinfo {pages}
  {7412--7424} (\bibinfo {year} {2017})}\BibitemShut {NoStop}%
\bibitem [{\citenamefont {Dima}\ and\ \citenamefont
  {Thirumalai}(2004)}]{pro2006}%
  \BibitemOpen
  \bibfield  {author} {\bibinfo {author} {\bibfnamefont {R.~I.}\ \bibnamefont
  {Dima}}\ and\ \bibinfo {author} {\bibfnamefont {D.}~\bibnamefont
  {Thirumalai}},\ }\bibfield  {title} {\enquote {\bibinfo {title} {Asymmetry in
  the shapes of folded and denatured states of proteins},}\ }\href {\doibase
  10.1021/jp037128y} {\bibfield  {journal} {\bibinfo  {journal} {The Journal of
  Physical Chemistry B}\ }\textbf {\bibinfo {volume} {108}},\ \bibinfo {pages}
  {6564--6570} (\bibinfo {year} {2004})}\BibitemShut {NoStop}%
\bibitem [{\citenamefont {Lobanov}, \citenamefont {Bogatyreva},\ and\
  \citenamefont {Galzitskaya}(2008)}]{rg2008}%
  \BibitemOpen
  \bibfield  {author} {\bibinfo {author} {\bibfnamefont {M.~Y.}\ \bibnamefont
  {Lobanov}}, \bibinfo {author} {\bibfnamefont {N.~S.}\ \bibnamefont
  {Bogatyreva}}, \ and\ \bibinfo {author} {\bibfnamefont {O.~V.}\ \bibnamefont
  {Galzitskaya}},\ }\bibfield  {title} {\enquote {\bibinfo {title} {Radius of
  gyration as an indicator of protein structure compactness},}\ }\href
  {\doibase 10.1134/S0026893308040195} {\bibfield  {journal} {\bibinfo
  {journal} {Molecular Biology}\ }\textbf {\bibinfo {volume} {42}},\ \bibinfo
  {pages} {623--628} (\bibinfo {year} {2008})}\BibitemShut {NoStop}%
\bibitem [{\citenamefont {Lee}\ and\ \citenamefont {Richards}(1971)}]{sasa71}%
  \BibitemOpen
  \bibfield  {author} {\bibinfo {author} {\bibfnamefont {B.}~\bibnamefont
  {Lee}}\ and\ \bibinfo {author} {\bibfnamefont {F.}~\bibnamefont {Richards}},\
  }\bibfield  {title} {\enquote {\bibinfo {title} {The interpretation of
  protein structures: Estimation of static accessibility},}\ }\href {\doibase
  https://doi.org/10.1016/0022-2836(71)90324-X} {\bibfield  {journal} {\bibinfo
   {journal} {Journal of Molecular Biology}\ }\textbf {\bibinfo {volume}
  {55}},\ \bibinfo {pages} {379 -- IN4} (\bibinfo {year} {1971})}\BibitemShut
  {NoStop}%
\bibitem [{\citenamefont {Horn}\ \emph {et~al.}(2012)\citenamefont {Horn},
  \citenamefont {Sengillo}, \citenamefont {Lin}, \citenamefont {Romo},\ and\
  \citenamefont {Grossfield}}]{Horn12}%
  \BibitemOpen
  \bibfield  {author} {\bibinfo {author} {\bibfnamefont {J.~N.}\ \bibnamefont
  {Horn}}, \bibinfo {author} {\bibfnamefont {J.~D.}\ \bibnamefont {Sengillo}},
  \bibinfo {author} {\bibfnamefont {D.}~\bibnamefont {Lin}}, \bibinfo {author}
  {\bibfnamefont {T.~D.}\ \bibnamefont {Romo}}, \ and\ \bibinfo {author}
  {\bibfnamefont {A.}~\bibnamefont {Grossfield}},\ }\bibfield  {title}
  {\enquote {\bibinfo {title} {Characterization of a potent antimicrobial
  lipopeptide via coarse-grained molecular dynamics},}\ }\href {\doibase
  https://doi.org/10.1016/j.bbamem.2011.07.025} {\bibfield  {journal} {\bibinfo
   {journal} {Biochimica et Biophysica Acta (BBA) - Biomembranes}\ }\textbf
  {\bibinfo {volume} {1818}},\ \bibinfo {pages} {212 -- 218} (\bibinfo {year}
  {2012})},\ \bibinfo {note} {membrane protein structure and
  function}\BibitemShut {NoStop}%
\bibitem [{\citenamefont {Tominaga}\ \emph {et~al.}(2010)\citenamefont
  {Tominaga}, \citenamefont {Mizuse}, \citenamefont {Hashidzume}, \citenamefont
  {Morishima},\ and\ \citenamefont {Sato}}]{Tominaga10}%
  \BibitemOpen
  \bibfield  {author} {\bibinfo {author} {\bibfnamefont {Y.}~\bibnamefont
  {Tominaga}}, \bibinfo {author} {\bibfnamefont {M.}~\bibnamefont {Mizuse}},
  \bibinfo {author} {\bibfnamefont {A.}~\bibnamefont {Hashidzume}}, \bibinfo
  {author} {\bibfnamefont {Y.}~\bibnamefont {Morishima}}, \ and\ \bibinfo
  {author} {\bibfnamefont {T.}~\bibnamefont {Sato}},\ }\bibfield  {title}
  {\enquote {\bibinfo {title} {Flower micelle of amphiphilic random copolymers
  in aqueous media},}\ }\href {\doibase 10.1021/jp104711q} {\bibfield
  {journal} {\bibinfo  {journal} {The Journal of Physical Chemistry B}\
  }\textbf {\bibinfo {volume} {114}},\ \bibinfo {pages} {11403--11408}
  (\bibinfo {year} {2010})}\BibitemShut {NoStop}%
\bibitem [{\citenamefont {Kuroda}, \citenamefont {Caputo},\ and\ \citenamefont
  {DeGrado}(2009)}]{Kuroda09}%
  \BibitemOpen
  \bibfield  {author} {\bibinfo {author} {\bibfnamefont {K.}~\bibnamefont
  {Kuroda}}, \bibinfo {author} {\bibfnamefont {G.~A.}\ \bibnamefont {Caputo}},
  \ and\ \bibinfo {author} {\bibfnamefont {W.~F.}\ \bibnamefont {DeGrado}},\
  }\bibfield  {title} {\enquote {\bibinfo {title} {The role of hydrophobicity
  in the antimicrobial and hemolytic activities of polymethacrylate
  derivatives},}\ }\href {\doibase 10.1002/chem.200801523} {\bibfield
  {journal} {\bibinfo  {journal} {Chem. Eur. J.}\ }\textbf {\bibinfo {volume}
  {15}},\ \bibinfo {pages} {1123--1133} (\bibinfo {year} {2009})}\BibitemShut
  {NoStop}%
\bibitem [{\citenamefont {Baul}\ and\ \citenamefont
  {Vemparala}(2017)}]{baul17}%
  \BibitemOpen
  \bibfield  {author} {\bibinfo {author} {\bibfnamefont {U.}~\bibnamefont
  {Baul}}\ and\ \bibinfo {author} {\bibfnamefont {S.}~\bibnamefont
  {Vemparala}},\ }\bibfield  {title} {\enquote {\bibinfo {title} {Influence of
  lipid composition of model membranes on methacrylate antimicrobial
  polymer–membrane interactions},}\ }\href {\doibase 10.1039/C7SM01211J}
  {\bibfield  {journal} {\bibinfo  {journal} {Soft Matter}\ }\textbf {\bibinfo
  {volume} {13}},\ \bibinfo {pages} {7665--7676} (\bibinfo {year}
  {2017})}\BibitemShut {NoStop}%
\bibitem [{\citenamefont {Sato}\ and\ \citenamefont {Feix}(2006)}]{Sato06}%
  \BibitemOpen
  \bibfield  {author} {\bibinfo {author} {\bibfnamefont {H.}~\bibnamefont
  {Sato}}\ and\ \bibinfo {author} {\bibfnamefont {J.~B.}\ \bibnamefont
  {Feix}},\ }\bibfield  {title} {\enquote {\bibinfo {title} {Peptide–membrane
  interactions and mechanisms of membrane destruction by amphipathic $\alpha-$
  helical antimicrobial peptides},}\ }\href {\doibase
  https://doi.org/10.1016/j.bbamem.2006.02.021} {\bibfield  {journal} {\bibinfo
   {journal} {Biochimica et Biophysica Acta (BBA) - Biomembranes}\ }\textbf
  {\bibinfo {volume} {1758}},\ \bibinfo {pages} {1245 -- 1256} (\bibinfo {year}
  {2006})},\ \bibinfo {note} {membrane Biophysics of Antimicrobial
  Peptides}\BibitemShut {NoStop}%
\bibitem [{\citenamefont {Pistolesi}, \citenamefont {Pogni},\ and\
  \citenamefont {Feix}(2007)}]{Pistolesi07}%
  \BibitemOpen
  \bibfield  {author} {\bibinfo {author} {\bibfnamefont {S.}~\bibnamefont
  {Pistolesi}}, \bibinfo {author} {\bibfnamefont {R.}~\bibnamefont {Pogni}}, \
  and\ \bibinfo {author} {\bibfnamefont {J.~B.}\ \bibnamefont {Feix}},\
  }\bibfield  {title} {\enquote {\bibinfo {title} {Membrane insertion and
  bilayer perturbation by antimicrobial peptide cm15},}\ }\href {\doibase
  https://doi.org/10.1529/biophysj.107.104034} {\bibfield  {journal} {\bibinfo
  {journal} {Biophysical Journal}\ }\textbf {\bibinfo {volume} {93}},\ \bibinfo
  {pages} {1651 -- 1660} (\bibinfo {year} {2007})}\BibitemShut {NoStop}%
\bibitem [{\citenamefont {Colak}\ \emph {et~al.}(2009)\citenamefont {Colak},
  \citenamefont {Nelson}, \citenamefont {N{\"u}sslein},\ and\ \citenamefont
  {Tew}}]{Colak09}%
  \BibitemOpen
  \bibfield  {author} {\bibinfo {author} {\bibfnamefont {S.}~\bibnamefont
  {Colak}}, \bibinfo {author} {\bibfnamefont {C.~F.}\ \bibnamefont {Nelson}},
  \bibinfo {author} {\bibfnamefont {K.}~\bibnamefont {N{\"u}sslein}}, \ and\
  \bibinfo {author} {\bibfnamefont {G.~N.}\ \bibnamefont {Tew}},\ }\bibfield
  {title} {\enquote {\bibinfo {title} {Hydrophilic modifications of an
  amphiphilic polynorbornene and the effects on its hemolytic and antibacterial
  activity},}\ }\href {\doibase 10.1021/bm801129y} {\bibfield  {journal}
  {\bibinfo  {journal} {Biomacromolecules}\ }\textbf {\bibinfo {volume} {10}},\
  \bibinfo {pages} {353--359} (\bibinfo {year} {2009})}\BibitemShut {NoStop}%
\bibitem [{\citenamefont {{\'A}lvarez-Paino}\ \emph {et~al.}(2015)\citenamefont
  {{\'A}lvarez-Paino}, \citenamefont {Mu{\~n}oz-Bonilla}, \citenamefont
  {L{\'o}pez-Fabal}, \citenamefont {G{\'o}mez-Garc{\'e}s}, \citenamefont
  {Heuts},\ and\ \citenamefont {Fern{\'a}ndez-Garc{\'i}a}}]{Alvarez15}%
  \BibitemOpen
  \bibfield  {author} {\bibinfo {author} {\bibfnamefont {M.}~\bibnamefont
  {{\'A}lvarez-Paino}}, \bibinfo {author} {\bibfnamefont {A.}~\bibnamefont
  {Mu{\~n}oz-Bonilla}}, \bibinfo {author} {\bibfnamefont {F.}~\bibnamefont
  {L{\'o}pez-Fabal}}, \bibinfo {author} {\bibfnamefont {J.}~\bibnamefont
  {G{\'o}mez-Garc{\'e}s}}, \bibinfo {author} {\bibfnamefont {J.~P.~A.}\
  \bibnamefont {Heuts}}, \ and\ \bibinfo {author} {\bibfnamefont
  {M.}~\bibnamefont {Fern{\'a}ndez-Garc{\'i}a}},\ }\bibfield  {title} {\enquote
  {\bibinfo {title} {Effect of glycounits on the antimicrobial properties and
  toxicity behavior of polymers based on quaternized dmaema},}\ }\href
  {\doibase 10.1021/bm5014876} {\bibfield  {journal} {\bibinfo  {journal}
  {Biomacromolecules}\ }\textbf {\bibinfo {volume} {16}},\ \bibinfo {pages}
  {295--303} (\bibinfo {year} {2015})}\BibitemShut {NoStop}%
\bibitem [{\citenamefont {Wong}\ \emph {et~al.}(2016)\citenamefont {Wong},
  \citenamefont {Khin}, \citenamefont {Ravikumar}, \citenamefont {Si},
  \citenamefont {Rice},\ and\ \citenamefont {Chan-Park}}]{Wong16}%
  \BibitemOpen
  \bibfield  {author} {\bibinfo {author} {\bibfnamefont {E.~H.~H.}\
  \bibnamefont {Wong}}, \bibinfo {author} {\bibfnamefont {M.~M.}\ \bibnamefont
  {Khin}}, \bibinfo {author} {\bibfnamefont {V.}~\bibnamefont {Ravikumar}},
  \bibinfo {author} {\bibfnamefont {Z.}~\bibnamefont {Si}}, \bibinfo {author}
  {\bibfnamefont {S.~A.}\ \bibnamefont {Rice}}, \ and\ \bibinfo {author}
  {\bibfnamefont {M.~B.}\ \bibnamefont {Chan-Park}},\ }\bibfield  {title}
  {\enquote {\bibinfo {title} {Modulating antimicrobial activity and mammalian
  cell biocompatibility with glucosamine-functionalized star polymers},}\
  }\href {\doibase 10.1021/acs.biomac.5b01766} {\bibfield  {journal} {\bibinfo
  {journal} {Biomacromolecules}\ }\textbf {\bibinfo {volume} {17}},\ \bibinfo
  {pages} {1170--1178} (\bibinfo {year} {2016})}\BibitemShut {NoStop}%
\end{thebibliography}%
\newpage
\onecolumngrid
\section*{\large Supplementary Information }
\setcounter{figure}{0}
\setcounter{table}{0}
\vspace*{10px}

\begin{figure*}[h]
 \centering
\includegraphics[width=6.5in]{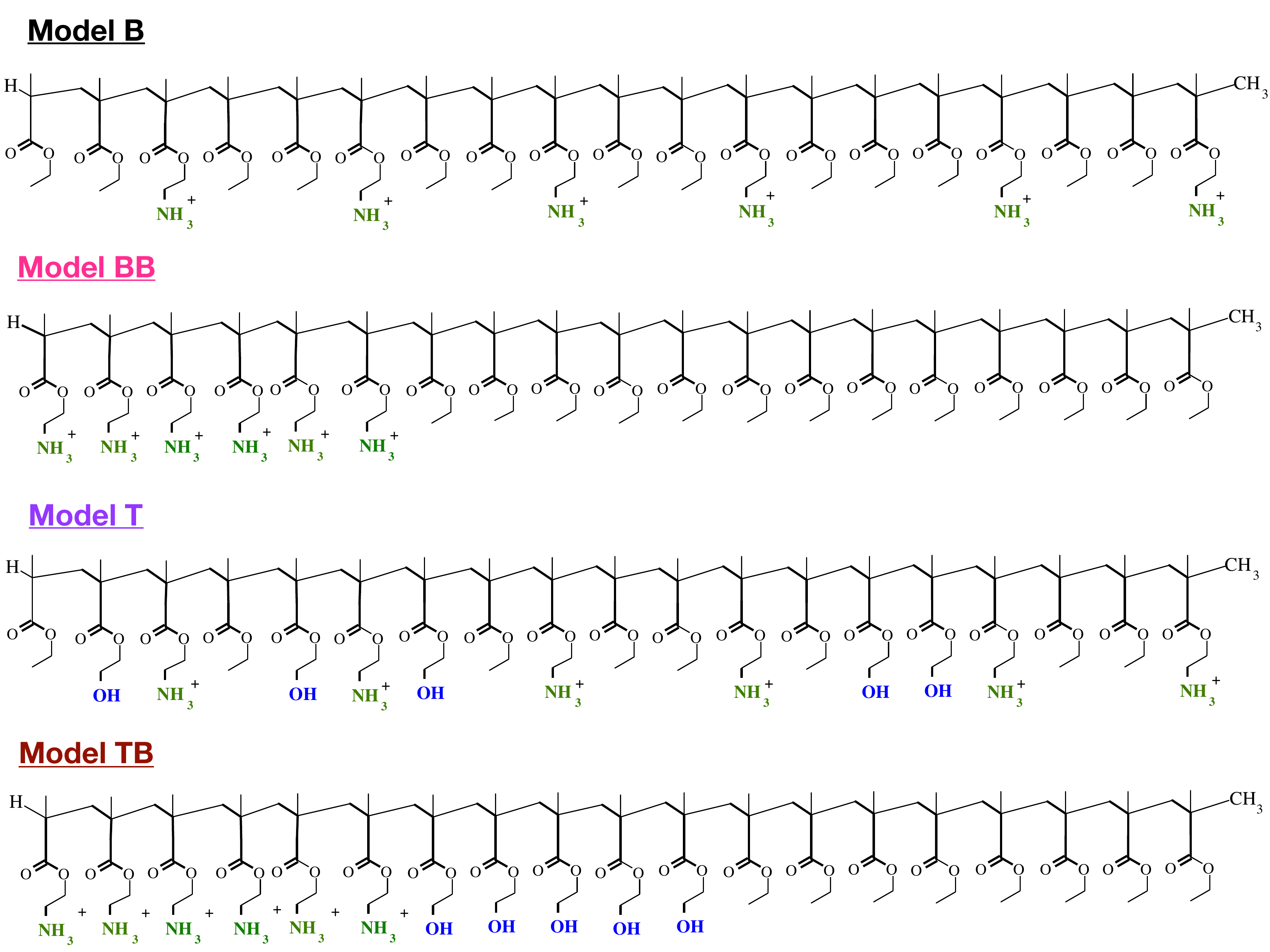}
\caption{\footnotesize Chemical structures of binary copolymers (models B, BB) and ternary copolymers (models T, TB). In all the model polymers, degree of polymerization (DP) = 19 and the number of cationic groups is fixed to be 6 per polymer.}
 \label{fig:polymers}
 \end{figure*}
 
 \begin{figure*}[h]
 \centering
\includegraphics[width=6.5in]{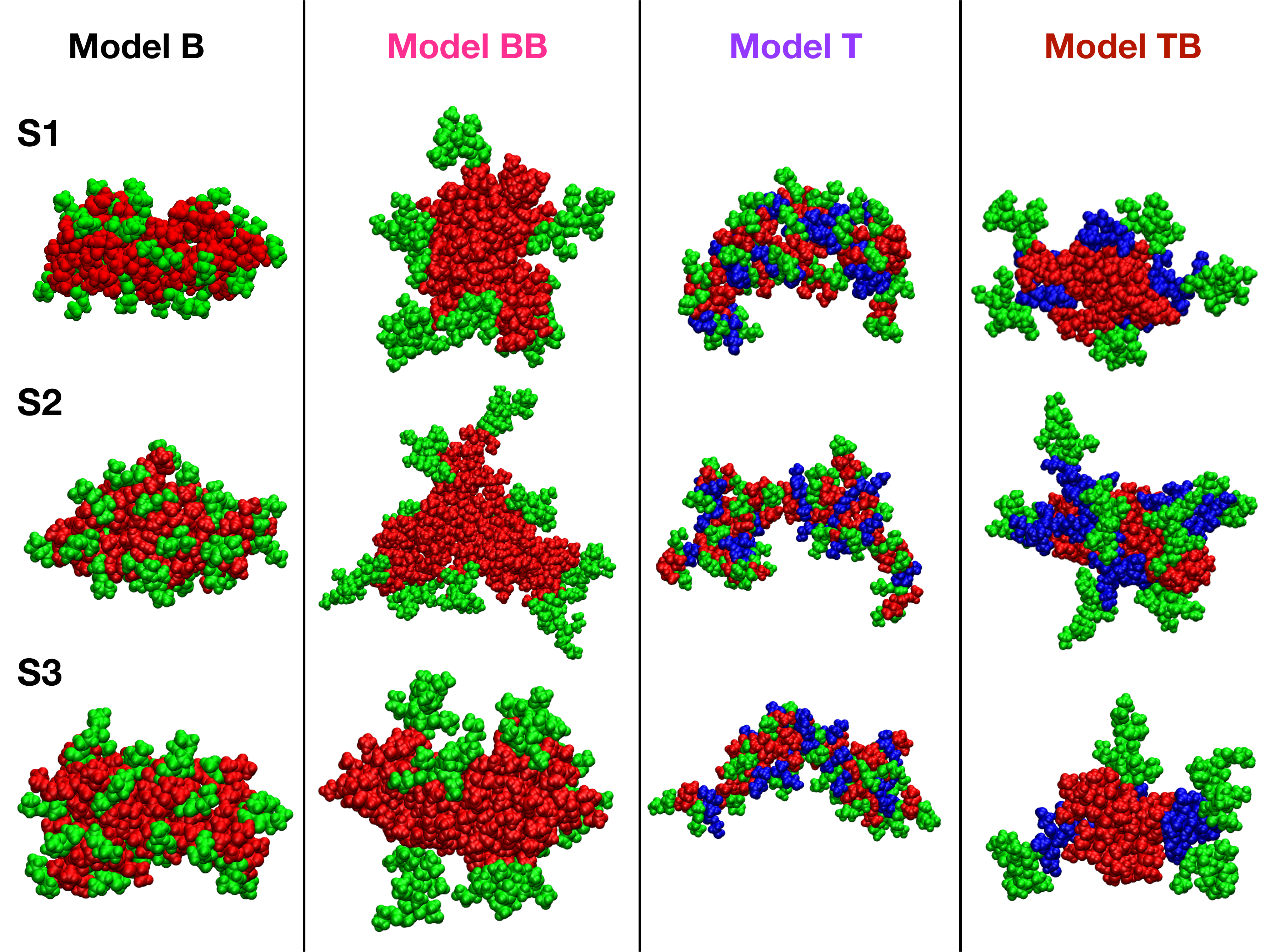}
\caption{\footnotesize The final conformations of the three independent simulations with hydrophobic groups (EMA, red); cationic side chain groups (AEMA, green); and polar side chain groups (HEMA, blue). Water molecules and ions are not shown for clarity.}
 \label{fig:agg}
 \end{figure*}
\newpage
\begin{table*}
\large
\centering
\begin{tabular}{|*{6}{c|}}\hline
\hline
\makebox[2.5em]{Model}& \makebox[2.5em]{Sequence} &\makebox[4em]{Aggr. run} &\makebox[6.5em]{Aggr. size, $N_{agg}$}\\
  &  & Time(ns) &   S1, S2, S3\\
\hline
   Binary copolymers  & (AEMA-HEMA)   & &  \\
  B & R  &  150, 150, 150 & 4, 4, 5 \\
 BB  &  B & 150, 150, 259 & 6, 10, 8 \\
 \hline
 Ternary copolymers & (AEMA-HEMA-EMA)  & &  \\
 T  &  R & 150, 150, 150 & 4, 5, 4  \\
 TB  & B & 150, 150, 150 & 5, 7, 4\\
\hline
\hline
\end{tabular}
\caption{Summary of copolymer systems considered and simulations performed. Here, R and B denote  random and block  sequence of the subunits. Three independent NPT runs  were performed for the aggregation of copolymers. Note that model BB (S3) has $N_{agg}$ = 5 till ~160 ns, after which it fuses with an aggregate of size, $N_{agg}$ = 3 to form bigger aggregate of size 8.}\label{tab:table1}
\end{table*}

 \begin{figure*}
 \centering
\includegraphics[width=8in]{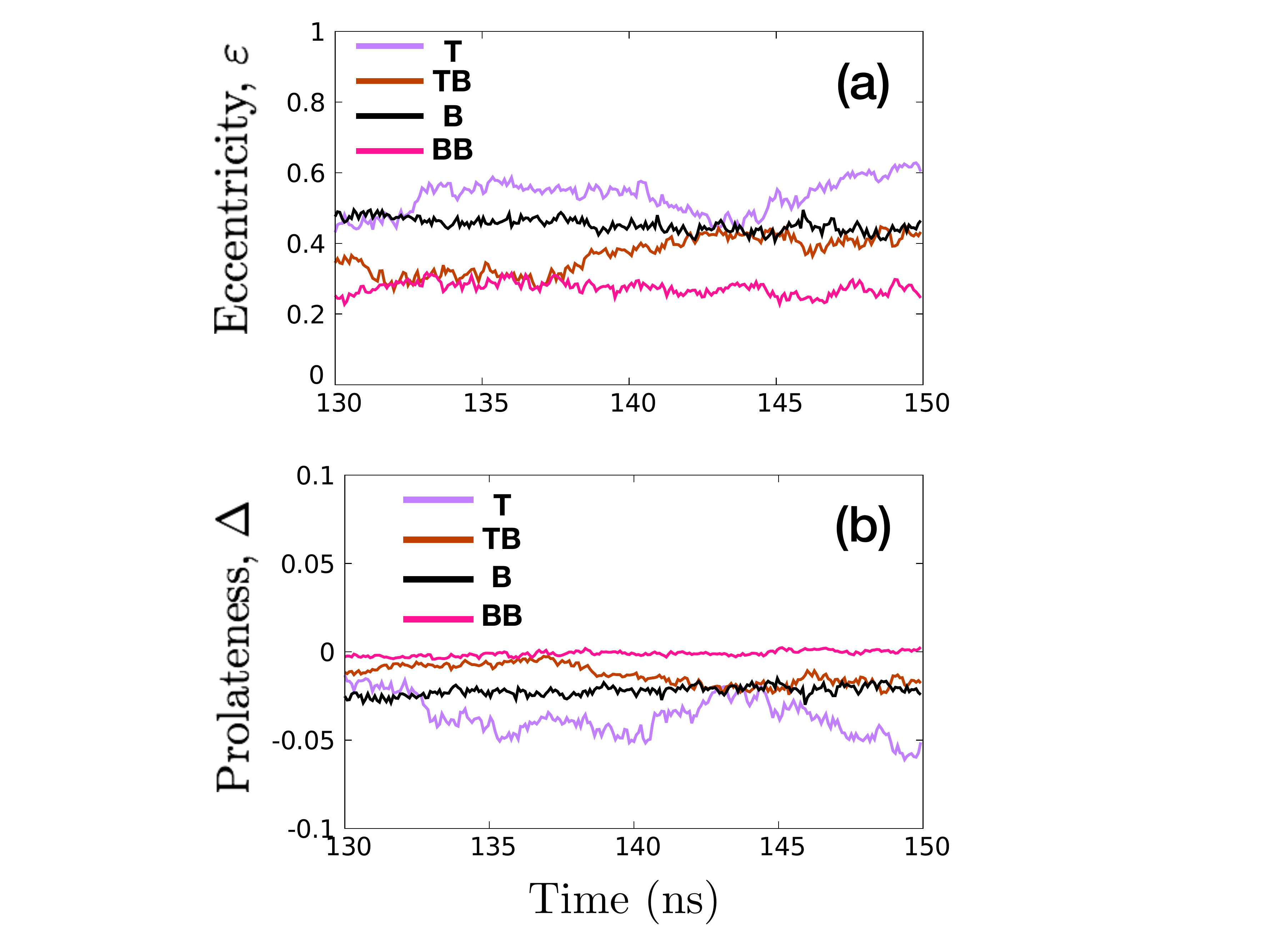}
\caption{ \footnotesize (a) Eccentricity, $\varepsilon$ and (b) Prolateness, $\Delta$ plotted as a function of simulation time from $130 -150$ ns  for all the model aggregates.}
\label{fig:ecctime}
\end{figure*}

 \begin{figure*}
 \centering
\includegraphics[width=6.5in]{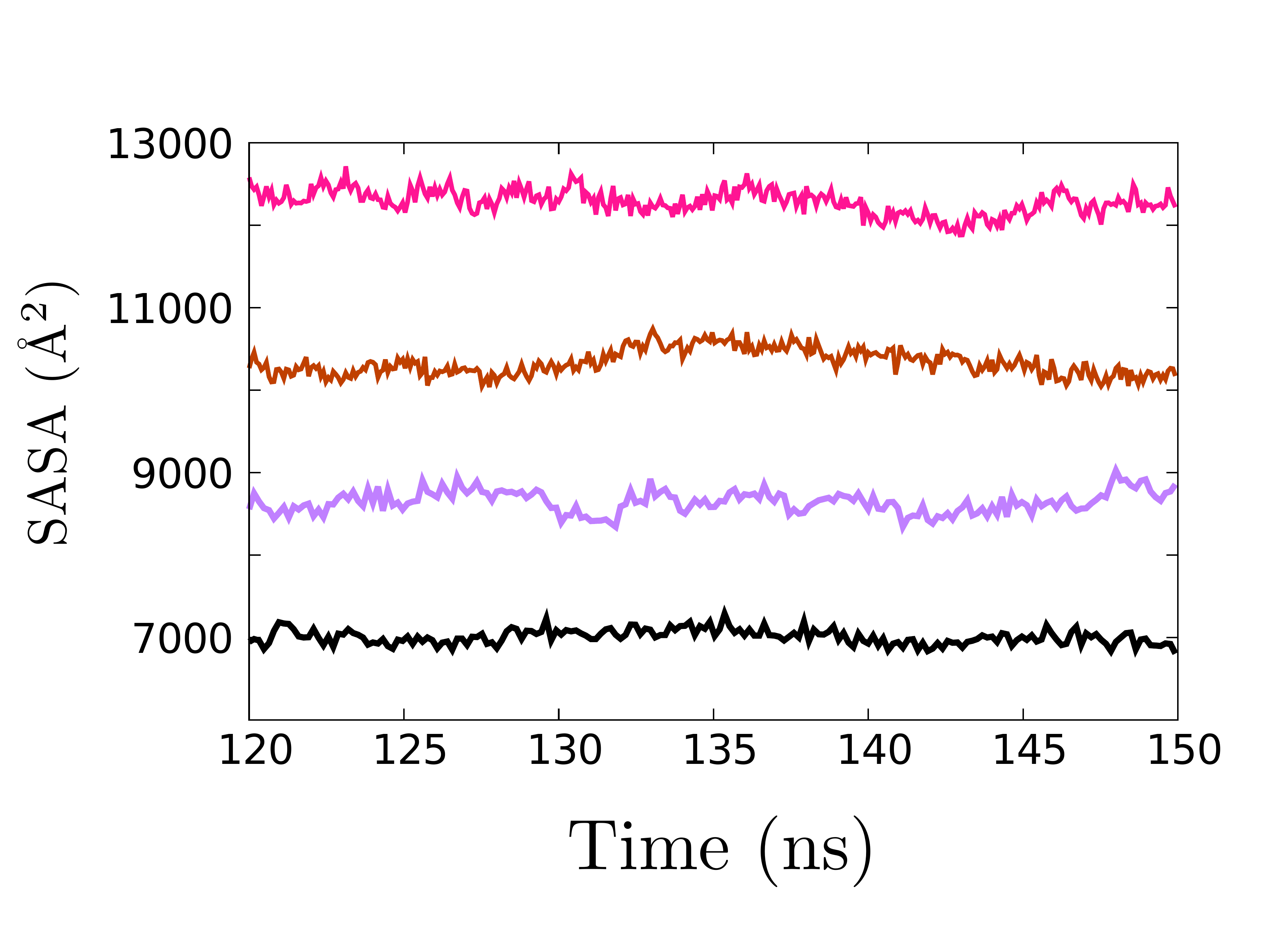}
\caption{ \footnotesize Total SASA values plotted as function as a function of simulation time for the model aggregates (from top to bottom - BB, TB, T, B).}
\label{fig:ecctime}
\end{figure*}

\begin{figure*}[h]
\centering
\includegraphics[width=6.5in]{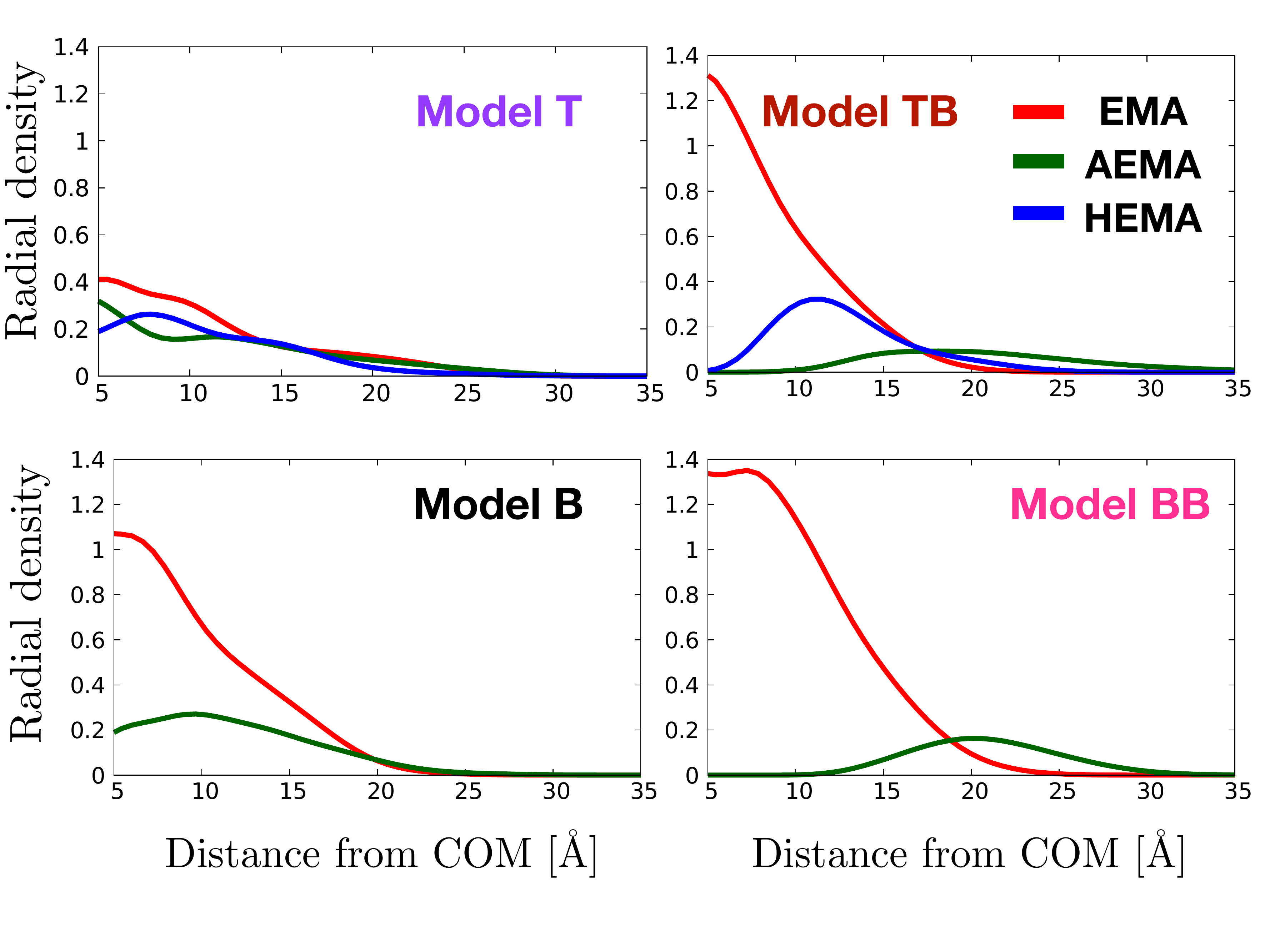}
\caption{\footnotesize (a) Radial density profiles of various components of the aggregates (AEMA, EMA and HEMA) for all the models considered in the study, calculated along the radial direction from the center of mass of the aggregate.}
\label{fig:radial-density}
\end{figure*}

\end{document}